\documentclass[
  twocolumn,
  prd,
  amssymb,
  preprintnumbers,
  nofootinbib,
  superscriptaddress,
  aps
]{revtex4-2}

\pdfoutput=1

\usepackage{graphicx}
\usepackage{amsmath,amssymb,amsfonts}
\usepackage{bm}
\usepackage{physics}        
\usepackage{siunitx}
\usepackage{booktabs}       
\usepackage{multirow}
\usepackage{dcolumn}        
\usepackage{enumitem}
\usepackage{microtype}      
\usepackage{placeins}      
\usepackage{xcolor}
\usepackage{xspace}   
\usepackage{float}
\usepackage{longtable}
\usepackage[space]{grffile}

\usepackage[caption=false]{subfig}

\definecolor{hypershade}{rgb}{0.18,0.18,0.6}
\usepackage{hyperref}
\hypersetup{
  colorlinks = true,
  linkcolor  = hypershade,
  citecolor  = {teal},
  urlcolor   = {purple},
  pdftitle   = {Sneezing Black hole},
  pdfauthor  = {Gaurav Rachh},
}

\usepackage[capitalize]{cleveref} 

\newcolumntype{d}[1]{D{.}{.}{#1}}

\begin{document}
    \title{Quasinormal Modes of Black Holes Embedded in Dark-Matter Halos: Analysis of Dehnen(1, 4, $\frac{\mathbf{5}}{\mathbf{2}}$) and DC14 Density Profiles}

\author{Gaurav Rachh}
\email{gauravphysics1212@gmail.com}
\affiliation{Master of Science, Department of Physics, Indian Institute of Science, C.~V.~Raman Avenue, Bengaluru 560012, India}

\begin{abstract}
We investigate how a surrounding dark-matter halo imprints itself on
the QNM spectrum of a static, spherically symmetric black hole,
considering two physically distinct density profiles: the Dehnen
$(1,4,5/2)$ and DC14 profiles. Both follow double-power-law forms,
while the shape parameters of the DC14 profile are calibrated to
galaxy-formation simulations and vary with the stellar-to-halo mass
ratio $X=\log_{10}(M_*/M_{\rm halo})$, which controls the inner density structure of the halo. For both profiles, we compute the halo contribution to the metric function and the resulting effective potential governing axial gravitational ($s=2$) perturbations, and extract the fundamental QNM frequencies for the $\ell=2$ multipole. Our results show that the halo scale radius $r_s$ has the largest influence on the QNM spectrum among the halo parameters explored, reshaping the effective potential barrier near the light ring and shifting both the real and imaginary parts of the QNM frequencies. The density normalization, by contrast, plays a comparatively minor role for the Dehnen profile but is more consequential for DC14, reflecting the different inner structures associated with its varying $X$ values. The Prony-extracted frequencies show excellent agreement with the sixth-order WKB results, with relative differences remaining below $0.04\%$ across the parameter ranges considered. To our knowledge, this constitutes the first QNM analysis of a black hole embedded in a DC14 dark-matter
halo. These results demonstrate the sensitivity of the QNM spectrum
to the surrounding dark-matter environment and provide a consistent numerical framework for studying such environmental effects.
\end{abstract}

\maketitle

\section{Introduction}
\label{sec:1}

General Relativity predicts that isolated black holes are among the
simplest compact objects, characterized by their mass, angular momentum, and electric charge. Astrophysical black holes,
however, are not expected to be perfectly isolated. They may reside in
complex environments containing accretion disks, stellar populations,
or dark-matter (DM) halos~~\cite{Barausse_2014,Jusufi_2020}. Such
environments modify the background spacetime and can consequently leave
an imprint on the gravitational-wave signal emitted by a perturbed black
hole.

The ringdown phase following the merger of compact objects is governed by the black hole perturbation theory~\cite{PhysRevD.110.124069}, which can be explained by the quasinormal modes (QNMs) of the remnant. These characteristic complex frequencies are determined by the
properties of the background spacetime and are independent of the details of the initial perturbation. QNMs, therefore, provide a natural framework for probing both the fundamental properties of black holes and their surrounding environments. The advent of gravitational-wave
astronomy~~\cite{Abbott2016GW150914,Aasi2015AdvancedLIGO,Acernese2015AdvancedVirgo,Akutsu2021KAGRA,LISA2017Science,ETScience2025} has made black hole spectroscopy an increasingly relevant observational prospect.

In black hole perturbation theory, the dynamics of linear perturbations are governed by wave equations derived by perturbing the field equations about a prescribed background geometry. For a broad class of static and spherically symmetric spacetimes, these equations
can be reduced to a Schr\"odinger-like master equation involving an effective potential~~\cite{PhysRevD.1.3514}. The QNM spectrum is then obtained by imposing physically appropriate boundary conditions, namely purely ingoing perturbations at the horizon and purely outgoing perturbations at spatial infinity. The resulting complex frequencies encode information about the geometry and, in the presence of matter, about the surrounding environment.

The theoretical study of QNMs has a long history, beginning with the analysis of black hole stability and scattering in the Schwarzschild spacetime~~\cite{PhysRev.108.1063,Zerilli1970, Vishveshwara1970, Goebel1972}. Subsequent developments established QNMs as a central component of gravitational-wave theory, with comprehensive reviews given in~~\cite{KokkotasSchmidt1999,BertiCardosoStarinets2009,
Konoplya_2011,Nollert1999}. The observation of binary black hole mergers has confirmed consistency of the post-merger remnant with the predictions of black hole perturbation theory~~\cite{Abbott2016ParameterEstimationGW150914}.

A particularly interesting environmental effect arises from dark matter. A dark-matter distribution surrounding a black hole modifies the metric, thereby changing the effective potential governing the perturbations. Even when the dark-matter density is sufficiently small that the geometry remains close to Schwarzschild, the resulting changes in the potential can produce measurable shifts in the QNM frequencies. Recent studies have investigated the influence of matter and dark-matter environments on black-hole ringdown, demonstrating that environmental effects can modify both the oscillation frequency and the damping time of the ringdown signal~\cite{Al_Badawi_2024, liang2025quasinormalmodesschwarzschildblack}. More recently, a systematic study of axial QNMs in generic matter halos showed that dilute environments produce a characteristic redshift of the QNM frequencies related to the compactness of the surrounding matter distribution~\cite{Pezzella_2025}.
Recently, Bolokhov~~\cite{bolokhov2025revisitingblackholesdarkmatter}
showed that the standard construction adopted in several recent
works imposing $g(r)=f(r)$ together with the Newtonian relation
between the enclosed mass and the tangential velocity generically fails to solve Einstein's field equations self-consistently for the matter source it purports to represent, and identified the
Dehnen-(1,4,5/2) profile as one such case~~\cite{Al_Badawi_2024,
liang2025quasinormalmodesschwarzschildblack}. Consistent with this,
we find that the lapse function reported in these works does not
approach unity as $r\to\infty$, in violation of asymptotic flatness
for a halo of finite total mass. In this work we therefore adopt the
Einstein-consistent Dehnen-(1,4,5/2) and DC14 metrics, and examine
the axial gravitational quasinormal-mode spectrum.

We adopt the approximation of neglecting perturbations of the dark-matter halo, i.e., $\delta T_{\mu\nu}=0$, treating it as a fixed background during ringdown, as commonly done in previous studies of axial gravitational perturbations in dark-matter environments ~\cite{Zhang_2021,Zhao2023DM,Zhao2024DM,liu2026blackholessurroundeddark}. The
axial gravitational perturbations are therefore governed by the
corresponding perturbed Einstein equations,
$\delta G_{\mu\nu}^{\rm (odd)}=0$.

We consider two physically distinct density profiles described by a generalized double-power-law form, with independently controllable scale radius and density normalization. The first is the Dehnen
$(1,4,5/2)$ profile, which has been widely used to model dark-matter halos, particularly in studies of dwarf galaxies. The second is the DC14 profile introduced by Di Cintio \textit{et al.}, which was constructed from cosmological hydrodynamical simulations and relates
the shape parameters of the halo to the stellar-to-halo mass ratio \cite{DiCintio2014}. In the DC14 model, the parameters $\alpha$, $\beta$, and $\gamma$ are not independent but are determined by
\begin{equation}
X = \log_{10}\left(\frac{M_*}{M_{\rm halo}}\right),
\label{eq:X_def}
\end{equation}
thereby providing an astrophysically motivated connection between the halo's inner structure and galaxy-formation processes.

In particular, the DC14 profile provides a useful framework for studying the transition between cuspy and cored dark-matter distributions associated with baryonic feedback, in which repeated, energetic gas outflows driven by \textit{supernovae} redistribute the central gravitational
potential and flatten an initially cuspy halo~\cite{DiCintio2014}. We investigate how this transition affects the black-hole QNM spectrum by
treating the characteristic scale radius $r_s$ and density normalization $\rho_s$ as phenomenological parameters in controlled scans, expressed in dimensionless form as $r_s/M$ and $\rho_s M^2$, following the
normalization commonly used in black-hole--dark-matter studies~\cite{jha2024shadowiscoquasinormalmodes}. For the DC14 profile, the shape parameters
$(\alpha,\beta,\gamma)$ are fixed by the stellar-to-halo mass ratio
$X$.

We additionally explore the dependence of the QNM spectrum on the stellar-to-halo mass ratio $X$, which controls the inner and outer slopes of the density distribution through the DC14 relations for $\alpha(X)$, $\beta(X)$, and $\gamma(X)$~~\cite{DiCintio2014}. At
fixed $X$, we independently vary $r_s$ and $\rho_s$ as phenomenological parameters in order to quantify the sensitivity of the QNM frequencies to the characteristic scale and density
normalization of the surrounding halo. The resulting QNM frequencies, extracted from the time-domain evolution via Prony analysis, are independently cross-checked against the sixth-order WKB
approximation, providing an additional validation of the numerical results.

The remainder of this paper is organized as follows. In Sec.~\ref{sec:dm_perturbations}, we develop the relativistic framework for a Schwarzschild black hole embedded in a dark-matter halo, deriving the background geometry and the axial gravitational perturbation equation. In Sec.~\ref{sec:dehnen_bolokhov}, we apply this framework to the Dehnen-(1,4,5/2) profile and discuss its Einstein-consistent metric. In Sec.~\ref{sec:DC14_dark-matter_halo}, we extend the analysis to the DC14 profile, with its shape parameters determined by the stellar-to-halo mass ratio $X$. Sec.~\ref{sec:Results} presents the fundamental $\ell=2$, $s=2$ QNM spectra obtained through time-domain evolution and Prony extraction, with independent comparison to the sixth-order WKB approximation using the implementation of Ref.~~\cite{Konoplya2026WKB}, together with validation against the Schwarzschild limit. Sec.~\ref{sec:conclusion} summarizes our findings and discusses their relevance for gravitational-wave spectroscopy of dark-matter environments.

\section{Schwarzschild Black Hole Surrounded by a dark-matter Halo}
\label{sec:dm_perturbations}

We consider a Schwarzschild black hole surrounded by a dark-matter
halo. Following Ref.~\cite{bolokhov2025revisitingblackholesdarkmatter},
we construct the background spacetime by solving the Einstein field
equations for the chosen dark-matter density profile. Imposing
$f(r)=g(r)$,
the resulting metric takes the form
\begin{equation}
ds^2
=
-f(r)\,dt^2
+\frac{dr^2}{f(r)}
+r^2d\theta^2
+r^2\sin^2\theta\,d\phi^2 .
\label{eq:spherical_metric}
\end{equation}

The corresponding energy-momentum tensor is
\begin{equation}
T^\mu_{\ \nu}
=
\mathrm{diag}
\left(
-\rho(r),
P_r(r),
P_t(r),
P_t(r)
\right),
\label{eq:energy_momentum}
\end{equation}
where $\rho$ is the energy density, $P_r$ is the radial pressure,
and $P_t$ is the tangential pressure. For the $f(r)=g(r)$
construction, the Einstein equations give
\begin{equation}
P_r(r)=-\rho(r),
\qquad
P_t(r)=-\frac{M_{\rm h}''(r)}{8\pi r},
\end{equation}
so that, in general, $P_t(r)\neq P_r(r)$.

Solving the Einstein field equations gives
\begin{equation}
g(r)-1+\frac{r g'(r)}{r^2}
=
8\pi T^t_{\ t}
=
-8\pi\rho(r),
\label{eq:einstein_tt}
\end{equation}

\begin{equation}
\frac{g(r)-1}{r^2}
+
\frac{g(r)}{f(r)}
\frac{f'(r)}{r}
=
8\pi T^r_{\ r}
=
8\pi P_r(r),
\label{eq:einstein_rr}
\end{equation}

and

\begin{equation}
\begin{aligned}
&\frac{g'(r)}{2}
\left(
\frac{1}{r}
+
\frac{f'(r)}{2f(r)}
\right)
\\
&\quad
+
\frac{g(r)}{2f(r)}
\left(
f''(r)
+
\frac{f'(r)}{r}
-
\frac{[f'(r)]^2}{2f(r)}
\right)
\\
&\qquad
= 8\pi T^\theta_{\ \theta}
= 8\pi T^\phi_{\ \phi}
= 8\pi P_t(r).
\end{aligned}
\label{eq:einstein_theta}
\end{equation}

where
\begin{equation}
f(r)
=
1-\frac{2M}{r}
-\frac{2M_{\rm h}(r)}{r},
\label{eq:metric_function}
\end{equation}
with $M$ denoting the black hole mass and $M_{\rm h}(r)$ the
enclosed mass of the dark-matter halo.

We now consider linear axial (odd-parity) gravitational perturbations
of this background. In the Regge--Wheeler gauge, the non-vanishing
components of the axial perturbation are
\begin{equation}
h_{tA}=h_0(t,r)S_A^{\ell m},
\qquad
h_{rA}=h_1(t,r)S_A^{\ell m},
\label{eq:axial_components}
\end{equation}
where $S_A^{\ell m}$ are the axial vector spherical harmonics. We
assume harmonic time dependence,
\begin{equation}
h_0(t,r)=h_0(r)e^{-i\omega t},
\qquad
h_1(t,r)=h_1(r)e^{-i\omega t}.
\label{eq:harmonic_perturbations}
\end{equation}

Substituting the perturbed metric into the Einstein field equations
and retaining terms to first order in the perturbation amplitude gives
three independent axial equations. The $t\phi$, $r\phi$, and
$\theta\phi$ components are
\begin{align}
{}&
r\left[-2M+r-2M_{\rm h}(r)\right]
\left[
i\omega\left(2h_1+r h_1'\right)
+r h_0''
\right]
\nonumber\\
&+
h_0
\left[
4M-r\ell(\ell+1)
+4M_{\rm h}(r)
+2r^2M_{\rm h}''(r)
\right]=0,
\label{eq:einstein_tphi}
\end{align}

\begin{equation}
\label{eq:einstein_rphi}
\begin{split}
&\frac{2M-r}{2M-r+2M_{\rm h}} \Bigg[ 2ir^2\omega h_0 - i r^3\omega h_0' \\
&\quad + h_1\Big\{ (2M-r)\big[\ell(\ell+1)-2\big] + r^3\omega^2 \\
&\quad + 2r(r-2M)M_{\rm h}'' \\
&\quad + 2M_{\rm h} \big[\ell(\ell+1)-2-2rM_{\rm h}''\big] \Big\} \Bigg] = 0.
\end{split}
\end{equation}

\begin{equation}
\label{eq:einstein_thetaphi}
\begin{split}
&\frac{i (2M - r) r^3 \omega h_0}{2M - r + 2M_{\rm h}} \\
&\quad + (2M - r) \bigg[ r (2M - r + 2M_{\rm h}) h_1' \\
&\qquad - 2h_1 \big(M + M_{\rm h} - r M_{\rm h}'\big) \bigg] = 0
\end{split}
\end{equation}

The $\theta\phi$ equation is algebraic in $h_0$ and can therefore be
used to eliminate $h_0$ in favour of $h_1$. Solving
Eq.~\eqref{eq:einstein_thetaphi}, we obtain
\begin{equation}
\begin{aligned}
h_0
={}&
\frac{i\left(2M-r+2M_{\rm h}(r)\right)}
{r^3\omega}
\Big[
r\left(2M-r+2M_{\rm h}(r)\right)h_1'
\\
&\qquad
-2h_1\left(
M+M_{\rm h}(r)-rM_{\rm h}'(r)
\right)
\Big].
\end{aligned}
\label{eq:h0_in_terms_h1}
\end{equation}

We next introduce the axial function $Z^{(-)}$ through
\begin{equation}
h_1(r)
=
\frac{r}{f(r)}Z^{(-)}(r).
\label{eq:master_variable}
\end{equation}

The tortoise coordinate is defined by
\begin{equation}
\frac{dr_*}{dr}
=
\frac{1}{f(r)},
\label{eq:tortoise_coordinate}
\end{equation}
which gives
\begin{equation}
\frac{d}{dr}
=
\frac{1}{f(r)}
\frac{d}{dr_*},
\label{eq:first_derivative_rstar}
\end{equation}
and
\begin{equation}
\frac{d^2Z^{(-)}}{dr^2}
=
\frac{1}{f(r)^2}
\frac{d^2Z^{(-)}}{dr_*^2}
-
\frac{f'(r)}{f(r)^2}
\frac{dZ^{(-)}}{dr_*}.
\label{eq:second_derivative_rstar}
\end{equation}

Substituting Eq.~\eqref{eq:h0_in_terms_h1} into the remaining
independent axial perturbation equation, together with the master
function defined in Eq.~\eqref{eq:master_variable}, and transforming
to the tortoise coordinate, we obtain the Schr\"odinger-like wave
equation

\begin{equation}
\frac{d^2Z^{(-)}}{dr_*^2}
+
\left[
\omega^2-V^{(-)}(r)
\right]Z^{(-)}
=0.
\label{eq:axial_wave_equation}
\end{equation}

The effective axial potential obtained from the linearized Einstein
equations is
\begin{equation}
\begin{aligned}
V^{(-)}(r)
={}&
\frac{2M+2M_{\rm h}(r)-r}{r^4}
\Big[
6M+6M_{\rm h}(r)
-r\ell(\ell+1)
\\
&\qquad
-2rM_{\rm h}'(r)
+2r^2M_{\rm h}''(r)
\Big].
\end{aligned}
\label{eq:axial_potential}
\end{equation}
Using
\footnotetext{We note that Eq.~\eqref{eq:axial_potential_compact} agrees with the
$F=G=f(r)$, $H=r^2$ specialization of the general axial effective
potential derived for a static, spherically symmetric spacetime of
the form $ds^2=-G(r)\,dt^2+F^{-1}(r)\,dr^2+H(r)\,d\Omega^2$ in
Ref.~~\cite{Zhang_2021}, providing an independent confirmation of our
result.}
\begin{equation}
2M+2M_{\rm h}(r)-r=-r f(r),
\label{eq:metric_identity}
\end{equation}
the effective potential can be written in the more convenient form
\begin{equation}
\begin{aligned}
V^{(-)}(r)
={}&
f(r)
\Bigg[
\frac{\ell(\ell+1)}{r^2}
-\frac{6\left[M+M_{\rm h}(r)\right]}{r^3}
\\
&\qquad
+\frac{2M_{\rm h}'(r)}{r^2}
-\frac{2M_{\rm h}''(r)}{r}
\Bigg].
\end{aligned}
\label{eq:axial_potential_compact}
\end{equation}

As a consistency check, when the halo contribution is removed,
$M_{\rm h}(r)\rightarrow0$, the metric function reduces to the
Schwarzschild form,
\begin{equation}
f(r)\rightarrow1-\frac{2M}{r},
\end{equation}
and Eq.~\eqref{eq:axial_potential_compact} becomes
\begin{equation}
V_{\rm RW}(r)
=
\left(1-\frac{2M}{r}\right)
\left[
\frac{\ell(\ell+1)}{r^2}
-\frac{6M}{r^3}
\right],
\label{eq:regge_wheeler_limit}
\end{equation}
which is the standard Regge--Wheeler potential for axial gravitational
perturbations of a Schwarzschild black hole.

\label{sec:2}


\section{Dehnen dark-matter Halo}
\label{sec:dehnen_bolokhov}
\subsection{Density Profile and Effective Mass}
We first explore the density profile of the Dehnen dark-matter halo. This halo is a specific example of a double-power-law profile ~\cite{Mo2010GalaxyFormation}. The corresponding density and enclosed mass profiles, together with the halo-induced shifts in the event horizon, photon sphere, and ISCO, are presented in Appendix~\ref{app:dehnen_properties}, where the corresponding numerical values
are also tabulated,
\begin{equation}
\rho_{\rm D}(r)
=
\frac{\rho_s}
{
\left(r/r_s\right)^\gamma
\left[
1+\left(r/r_s\right)^\alpha
\right]^{(\beta-\gamma)/\alpha}
}.
\label{eq:dehnen_general}
\end{equation}

Here, $\rho_s$ is the density normalization, $r_s$ is the
characteristic scale radius, and $(\alpha,\beta,\gamma)$ determine the
inner slope, outer slope, and sharpness of the transition between the
two regimes.

The standard Dehnen family has the asymptotic behaviors
\begin{equation}
\rho_{\rm D}(r)\propto r^{-\gamma},
\qquad
r\ll r_s,
\end{equation}
and
\begin{equation}
\rho_{\rm D}(r)\propto r^{-\beta},
\qquad
r\gg r_s.
\end{equation}

For the Dehnen model considered in this work, we adopt
\begin{equation}
(\alpha,\beta,\gamma)
=
\left(
1,4,\frac{5}{2}
\right).
\label{eq:dehnen_1452}
\end{equation}

The enclosed dark-matter mass is obtained from
\begin{equation}
M_{\rm D}(r)
=
4\pi
\int_0^r
\rho_{\rm D}(\tilde r)
\tilde r^2\,d\tilde r.
\label{eq:dehnen_mass_integral}
\end{equation}

Introducing
\begin{equation}
x=\frac{r}{r_s},
\end{equation}
the enclosed mass can be expressed in terms of the hypergeometric
function as

\begin{equation}
\label{eq:dehnen_mass}
\begin{split}
M_{\rm D}(r) &= \frac{4\pi\rho_s r_s^3}{3-\gamma}\, x^{3-\gamma} \\
&\quad \times {}_2F_1\bigg( \frac{3-\gamma}{\alpha}, \frac{\beta-\gamma}{\alpha}; 1+\frac{3-\gamma}{\alpha}, -x^\alpha \bigg)
\end{split}
\end{equation}

This mass profile is then inserted into Eq.~\eqref{eq:axial_potential_compact} to
construct the corresponding black hole effective potential.

\begin{figure}[htbp]
\centering
\includegraphics[width=1\linewidth]{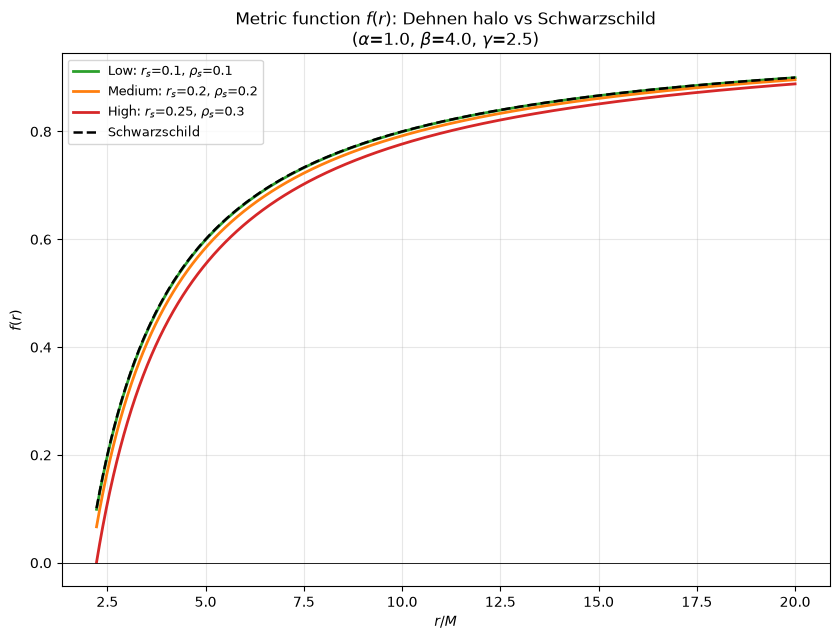}
\caption{
Metric function $f(r)$ for a Schwarzschild black hole surrounded by a
Dehnen dark-matter halo, for low, intermediate, and high
$(r_s,\rho_s)$ parameter combinations, compared against the vacuum
Schwarzschild case (dashed black). The additional enclosed halo mass
modifies the metric function and shifts the horizon's location.}
\label{fig:fr_dehnen}
\end{figure}

For illustration, Fig.~\ref{fig:fr_dehnen} shows the resulting metric
function $f(r)$ for three representative combinations of the halo
parameters: a low-parameter case
$(r_s=0.10,\rho_s=0.10)$, an intermediate case
$(r_s=0.20,\rho_s=0.20)$, and a high-parameter case
$(r_s=0.25,\rho_s=0.30)$, compared against the vacuum Schwarzschild
case.

The metric function for the Schwarzschild--Dehnen $(1,4,5/2)$
spacetime is
\begin{equation}
f(r)=g(r)
=
1-\frac{2M}{r}
-\frac{16\pi\rho_s r_s^3}
{\sqrt{r(r+r_s)}}.
\label{eq:schwarzschild_dehnen_metric}
\end{equation}

This solution is asymptotically flat and approaches the Schwarzschild
spacetime in the limit
\begin{equation}
\rho_s,r_s\rightarrow0.
\end{equation}
Importantly, this solution exactly satisfies the Einstein field
equations, as discussed by Bolokhov~\cite{bolokhov2025revisitingblackholesdarkmatter}.

The metric function is shown in Fig.~\ref{fig:fr_dehnen}. All curves
approach the Minkowski value,
\begin{equation}
f(r)\rightarrow1,
\qquad r\rightarrow\infty,
\end{equation}
while they exhibit the familiar Schwarzschild behavior
\begin{equation}
f(r)\sim-\frac{2M}{r},
\qquad r\rightarrow0.
\end{equation}
Thus, the dark-matter halo contribution primarily modifies the metric at intermediate and large radial distances.

\subsection{Effective Potential}

Having determined the enclosed dark-matter mass Eq.~\eqref{eq:dehnen_mass}, the metric function can be written as
\begin{equation}
f(r)
=
1-\frac{2M}{r}
-\frac{2M_{\rm D}(r)}{r}
=
1-\frac{2\left[M+M_{\rm D}(r)\right]}{r}.
\label{eq:f_mass}
\end{equation}
where $M_{\rm D}(r)\equiv M_{\rm Dehnen}(r)$ denotes the enclosed Dehnen dark-matter mass. The derivatives of the enclosed dark-matter mass can be evaluated directly
from the density profile,
\begin{equation}
M_{\rm D}'(r)=4\pi r^2\rho(r),
\label{eq:mass_first_derivative}
\end{equation}
while
\begin{equation}
M_{\rm D}''(r)
=
8\pi r\rho(r)+4\pi r^2\rho'(r).
\label{eq:mass_second_derivative}
\end{equation}

\begin{figure}[htbp]
    \centering
    \includegraphics[width=\linewidth]{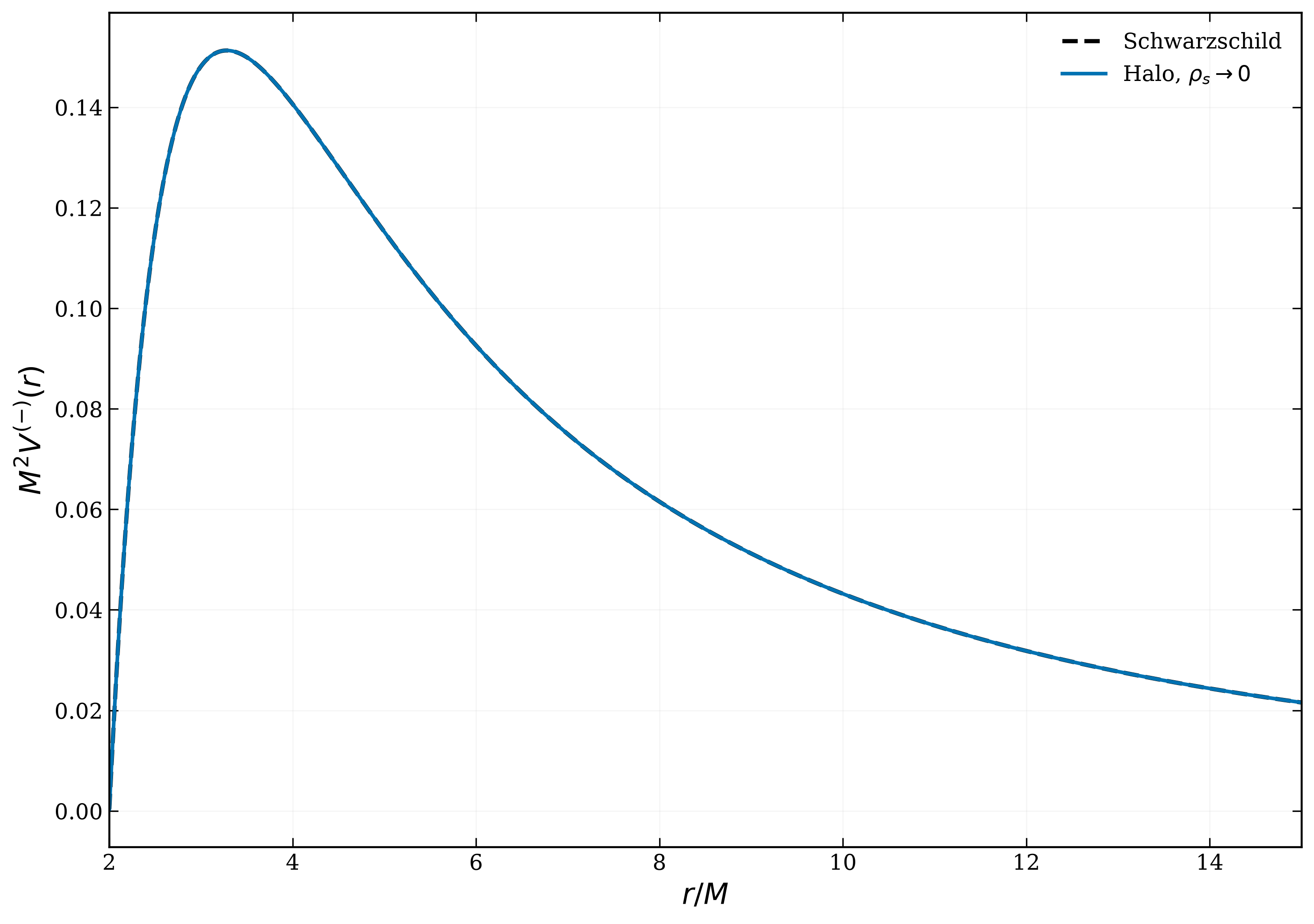}
    \caption{Effective potential for the Schwarzschild--Dehnen $(1,4,5/2)$ spacetime approaching the Schwarzschild limit.}
    \label{fig:potential_schw_halo_exact}
\end{figure}

As shown in Eq.~\eqref{eq:regge_wheeler_limit}, in the limit
$\rho_s\rightarrow0$, the effective potential reduces to the standard Regge--Wheeler potential.

\begin{figure}[htbp]
    \centering

    \includegraphics[width=\linewidth]{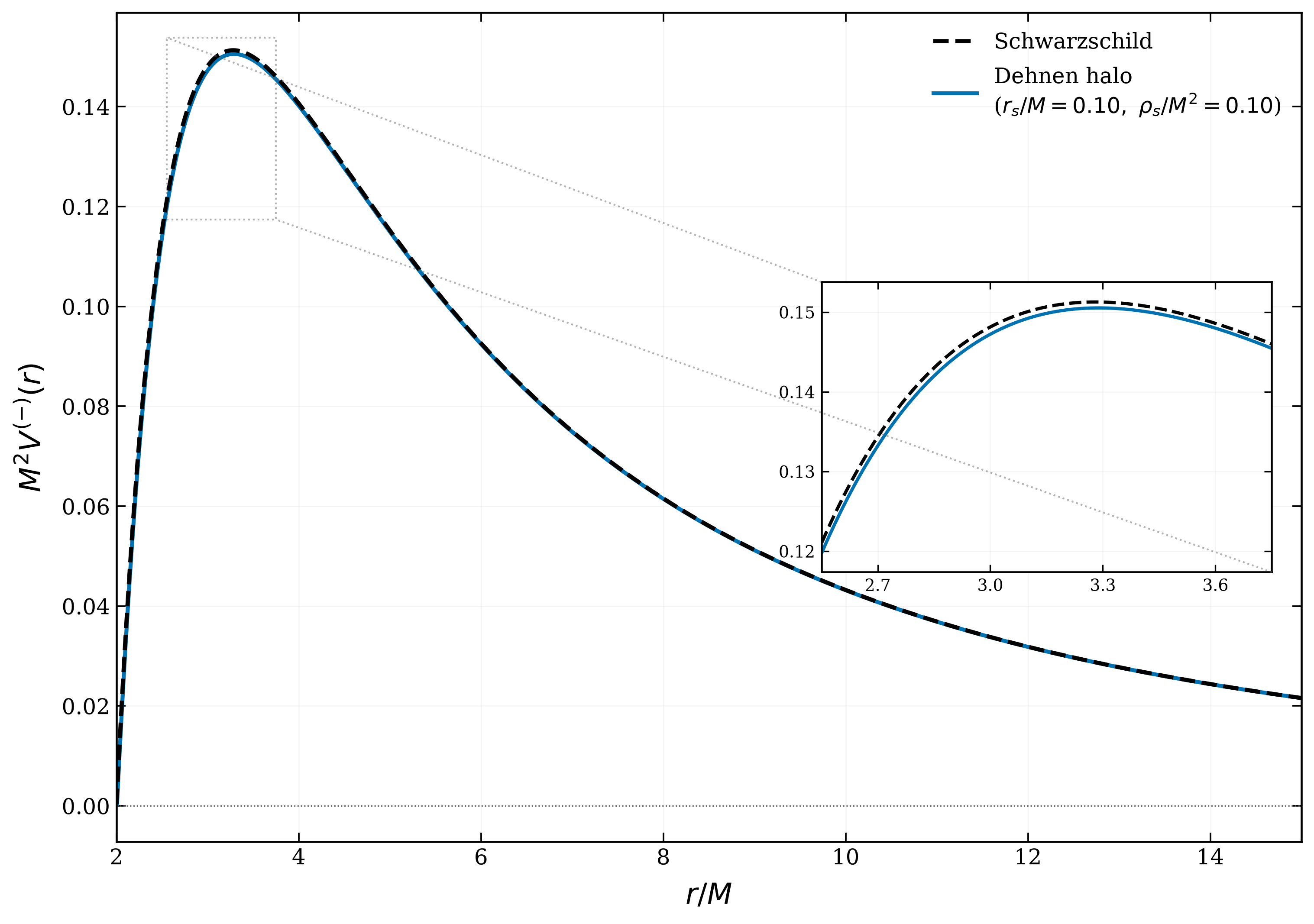}

    \caption{
    Axial gravitational effective potential
    $V^{(-)}(r)$ for a Schwarzschild black hole surrounded by
    a Dehnen dark-matter halo, compared with the corresponding
    vacuum Schwarzschild potential. The halo modifies both the
    height and radial location of the potential barrier.
    }
    \label{fig:potential_schw_halo}
\end{figure}

Figure~\ref{fig:potential_schw_halo} shows the effective potential $V^{(-)}(r)$ for the axial gravitational perturbations of the Schwarzschild black hole surrounded by the Dehnen dark-matter halo($\rho_s=0.1$ and $r_s=0.1$). For comparison, the vacuum Schwarzschild potential is also shown. The presence of the halo modifies the shape of the potential barrier, resulting in a shift of both its peak location and height. These changes in the effective potential provide the direct mechanism through which the dark-matter environment affects the quasinormal-mode spectrum. We next investigate the effects of the halo parameters separately by varying $\rho_s$ while keeping $r_s$ fixed, and then varying $r_s$ while keeping $\rho_s$ fixed.
\begin{figure}[htbp]
    \centering
    \includegraphics[width=0.96\linewidth]{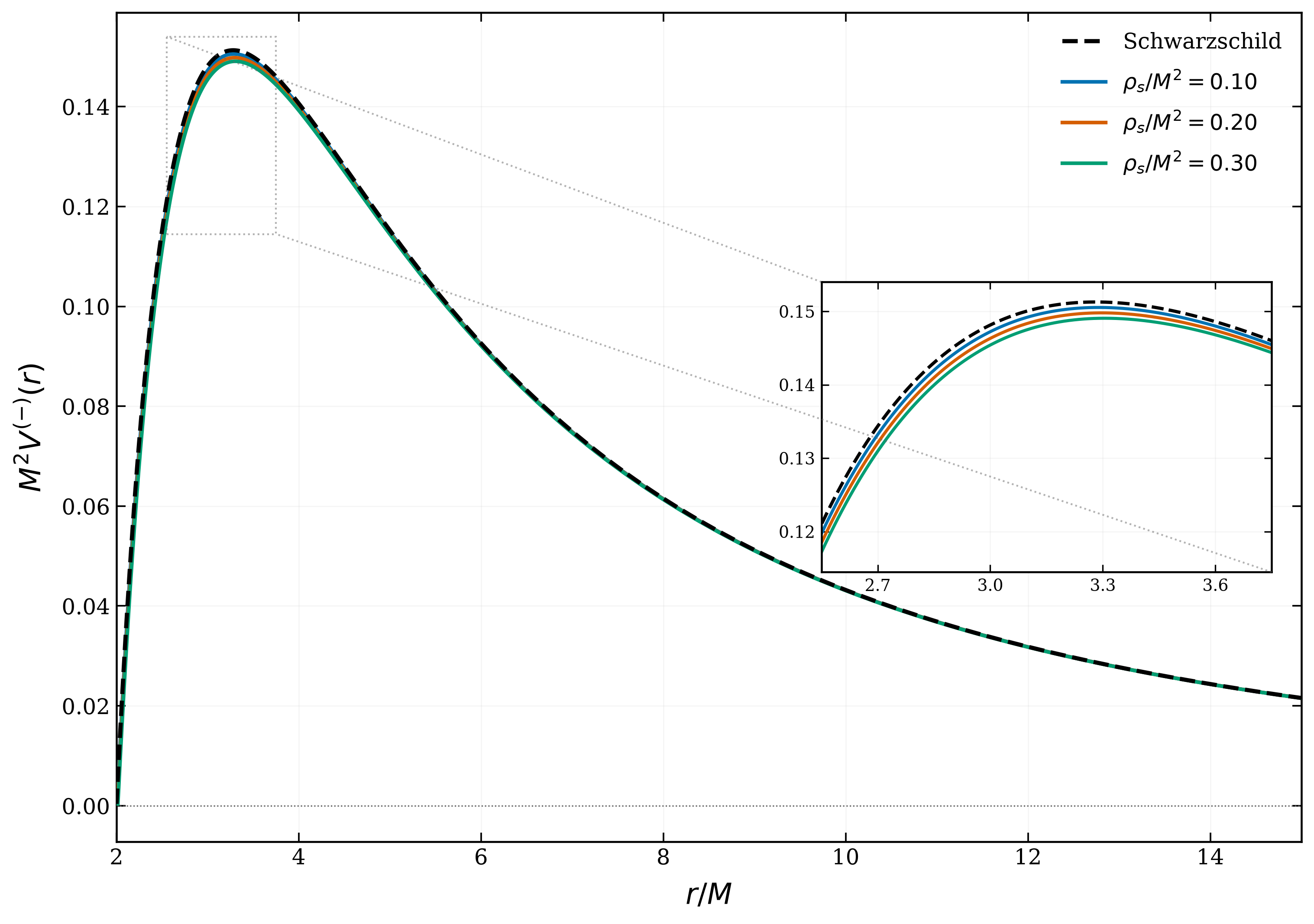}
    \caption{Dependence of the axial gravitational effective potential $V^{(-)}(r)$ on the parameters of the Dehnen dark-matter halo for $\ell=2$. This plot shows the effect of varying the density normalization $\rho_s$ at a fixed scale radius $r_s$.}
    \label{fig:potential_dehnen_rhos}
\end{figure}

\begin{figure}[htbp]
    \centering
    \includegraphics[width=0.96\linewidth]{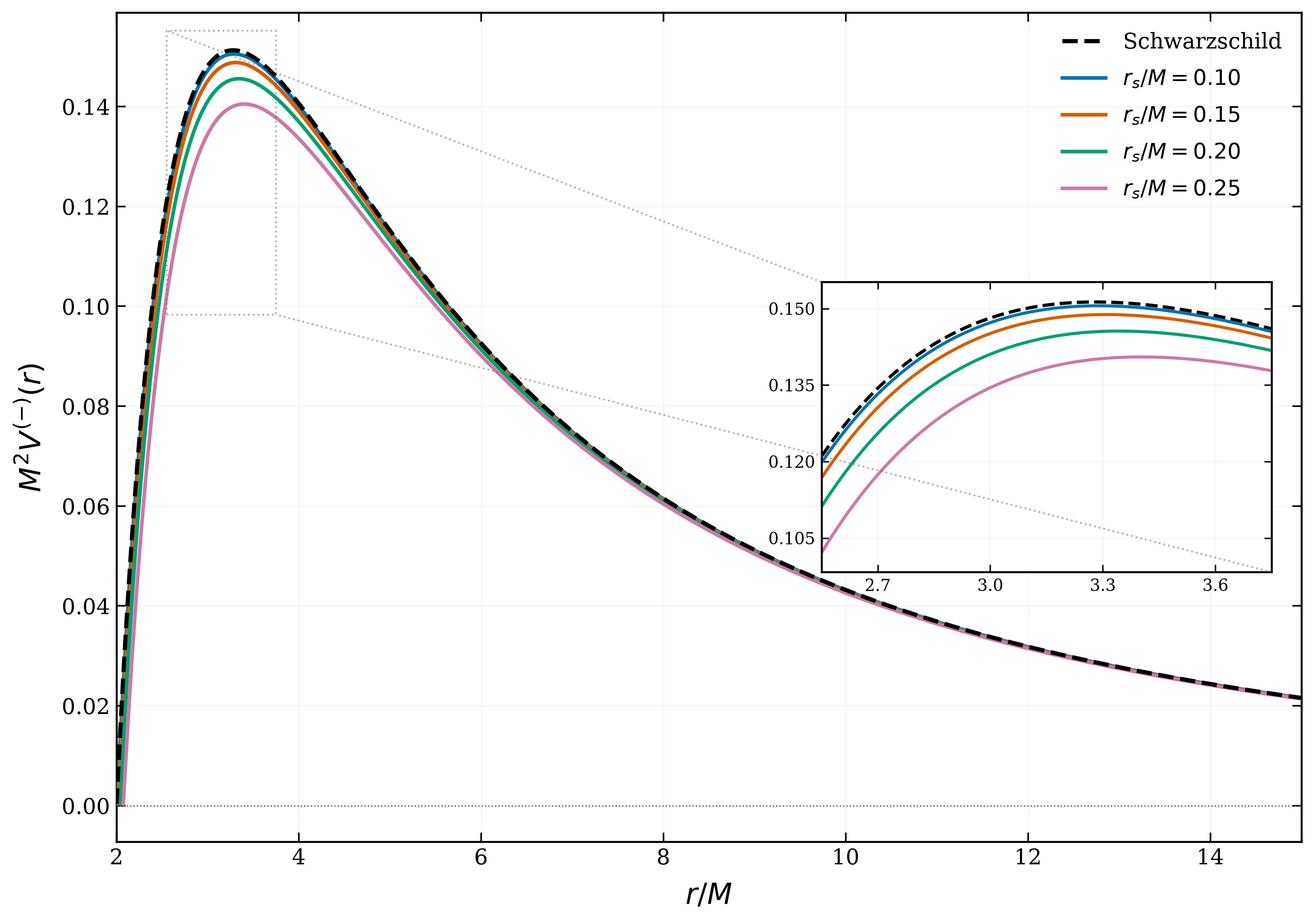}
    \caption{Dependence of the axial gravitational effective potential $V^{(-)}(r)$ on the parameters of the Dehnen dark-matter halo for $\ell=2$. This plot shows the effect of varying the scale radius $r_s$ at a fixed density normalization $\rho_s$.}
    \label{fig:potential_dehnen_rs}
\end{figure}

\begin{figure}[htbp]
    \centering
    \includegraphics[width=\linewidth]{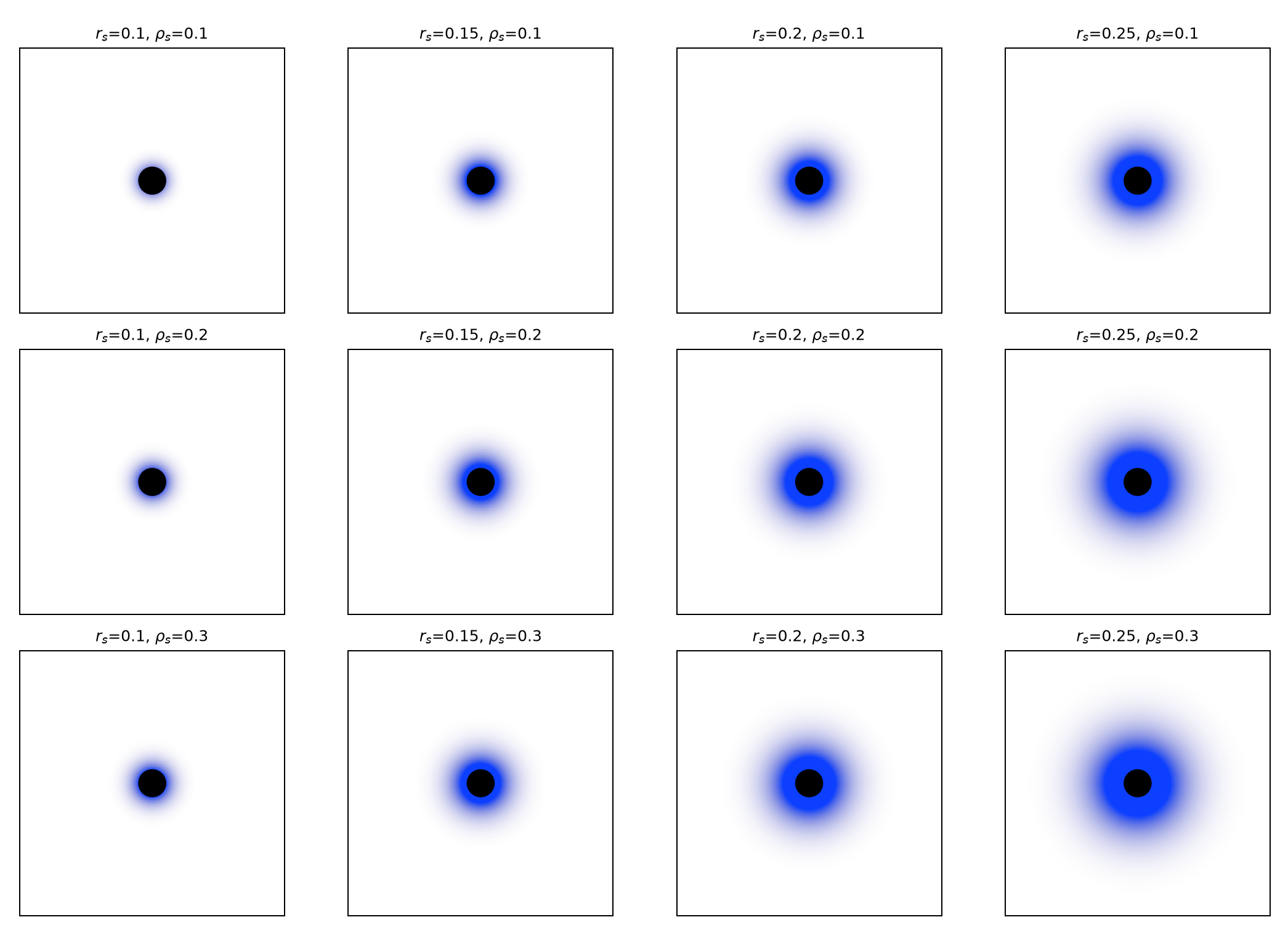}
    \caption{
    Illustrative density distribution of a Schwarzschild black hole
    surrounded by a Dehnen dark-matter halo. The panels show the
    variation of the halo structure with the scale radius $r_s$ and
    density normalization $\rho_s$. The columns correspond to
    $r_s=0.10,\ 0.15,\ 0.20,\ 0.25$, while the rows correspond to
    $\rho_s=0.10,\ 0.20,\ 0.30$, respectively. The black central
    region represents the black hole, while the surrounding blue
    distribution represents the dark-matter halo. Increasing $r_s$
    produces a more spatially extended halo, whereas increasing
    $\rho_s$ enhances the central density for fixed $r_s$.
    }
    \label{fig:dehnen_halo_parameter_scan}
\end{figure}

The dependence of the potential on the halo parameters is illustrated
in Figs.~\ref{fig:potential_dehnen_rhos} and
\ref{fig:potential_dehnen_rs}. We observe systematic changes in the
effective potential as $\rho_s$ and $r_s$ are varied. The resulting QNM
shifts are examined separately for the two halo parameters in the
following analysis.

\section{DC14 dark-matter Halo}
\label{sec:DC14_dark-matter_halo}

\subsection{Density Profile and Effective Mass}

We next consider the DC14 dark-matter profile. Its density is written in the same generalized form,
\begin{equation}
\rho_{\rm DC14}(r)
=
\frac{\rho_s}
{
\left(r/r_s\right)^\gamma
\left[
1+\left(r/r_s\right)^\alpha
\right]^{(\beta-\gamma)/\alpha}
}.
\label{eq:dc14_density}
\end{equation}
The corresponding density and enclosed mass profiles, along with the resulting shifts in the event-horizon, photon-sphere, and ISCO radii, are presented in Appendix~\ref{app:dc14_properties}. The appendix also summarizes the dependence of the density and enclosed mass on the DC14 parameters and provides
the corresponding numerical values. 

The DC14 profile was introduced by Di Cintio et al.~~\cite{DiCintio2014}
as a special case of the general and flexible
$(\alpha,\beta,\gamma)$ profile
~\cite{Jaffe1983,Hernquist1990,Zhao1996}. In the
DC14 model, the shape parameters $(\alpha,\beta,\gamma)$ are not
independent. Instead, they are determined by the stellar-to-halo mass
ratio
\begin{equation}
X=
\log_{10}
\left(
\frac{M_*}{M_{\rm halo}}
\right).
\label{eq:dc14_X}
\end{equation}
Defining
\begin{equation}
A=10^{X+2.33},
\qquad
B=10^{X+2.56},
\end{equation}
The three shape parameters are
\begin{align}
\alpha(X)
&=
2.94
-
\log_{10}
\left[
A^{-1.08}
+
A^{2.29}
\right],
\label{eq:dc14_alpha}
\\
\beta(X)
&=
4.23
+
1.34X
+
0.26X^2,
\label{eq:dc14_beta}
\\
\gamma(X)
&=
-0.06
+
\log_{10}
\left[
B^{-0.68}
+
B
\right].
\label{eq:dc14_gamma}
\end{align}
The relations are calibrated over approximately
\begin{equation}
-4.1 < X < -1.3,
\end{equation}
the range explored by the underlying simulations
~\cite{DiCintio2014}.
An important feature of the DC14 prescription is the non-monotonic
behavior of the inner slope $\gamma$. The minimum value of $\gamma$
occurs over the interval
\begin{equation}
-2.6 \lesssim X \lesssim -2.4,
\end{equation}
corresponding to the strongest flattening of the inner dark-matter
profile in the DC14 calibration
~\cite{DiCintio2014}. This makes $X\simeq-2.6$ particularly useful for
studying the influence of a cored or weakly cusped environment on
black hole perturbations.

In the parameter study, the parameter $X$ determines the shape
parameters $(\alpha,\beta,\gamma)$ through Eqs.~\eqref{eq:dc14_alpha}--\eqref{eq:dc14_gamma}, while the scale
radius $r_s$ and density normalization $\rho_s$ are varied phenomenologically to investigate their influence on the effective potential and, subsequently, on the QNM spectrum. We emphasize that this differs from the treatment in the original DC14 calibration, where $r_s$ and $\rho_s$ are instead fixed by matching the total halo mass at the virial radius together with a concentration--mass relation~~\cite{DiCintio2014}. Here we treat $r_s$ and $\rho_s$ as
independent phenomenological parameters, allowing a controlled scan of their influence on the QNM spectrum at fixed $X$.

\begin{figure}[H]
    \centering
    \includegraphics[width=\columnwidth]{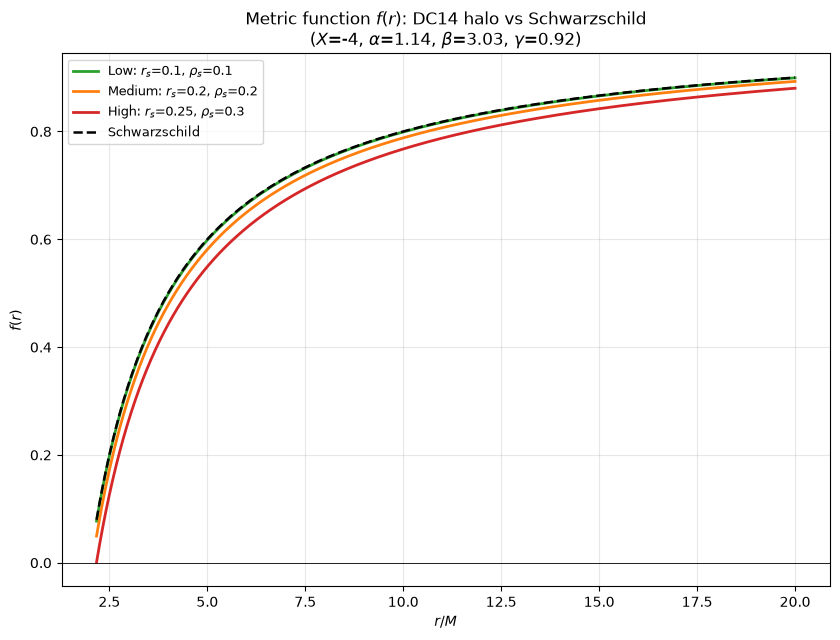}
    \caption{Metric function $f(r)$ for a Schwarzschild black hole surrounded by a DC14 dark-matter halo, shown for low, intermediate, and high $(r_s,\rho_s)$ combinations at a stellar-to-halo mass ratio of $X=-4$.}
    \label{fig:fr_dc14_Xm4}
\end{figure}

\begin{figure}[H]
    \centering
    \includegraphics[width=0.7\linewidth]{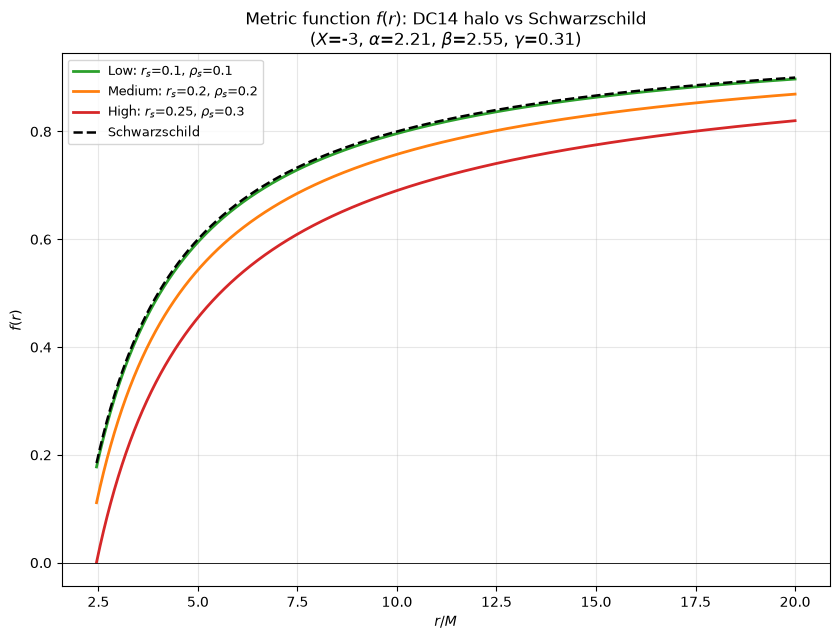}
    \caption{Metric function $f(r)$ for a Schwarzschild black hole surrounded by a DC14 dark-matter halo for a stellar-to-halo mass ratio of $X=-3$.}
    \label{fig:fr_dc14_Xm3}
\end{figure}

\begin{figure}[H]
    \centering
    \includegraphics[width=0.7\linewidth]{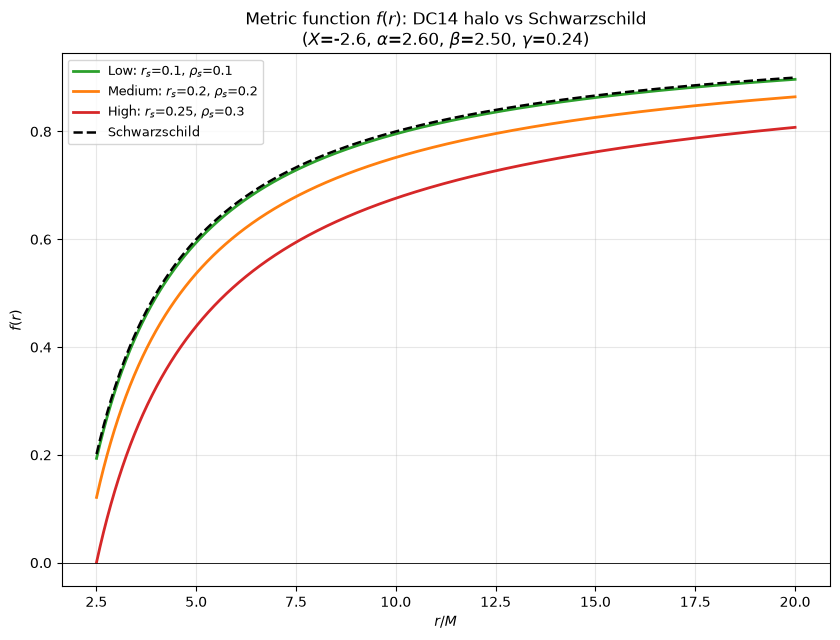}
    \caption{Metric function $f(r)$ for a Schwarzschild black hole surrounded by a DC14 dark-matter halo for a stellar-to-halo mass ratio of $X=-2.6$.}
    \label{fig:fr_dc14_Xm2p6}
\end{figure}

\begin{figure}[H]
    \centering
    \includegraphics[width=0.7\linewidth]{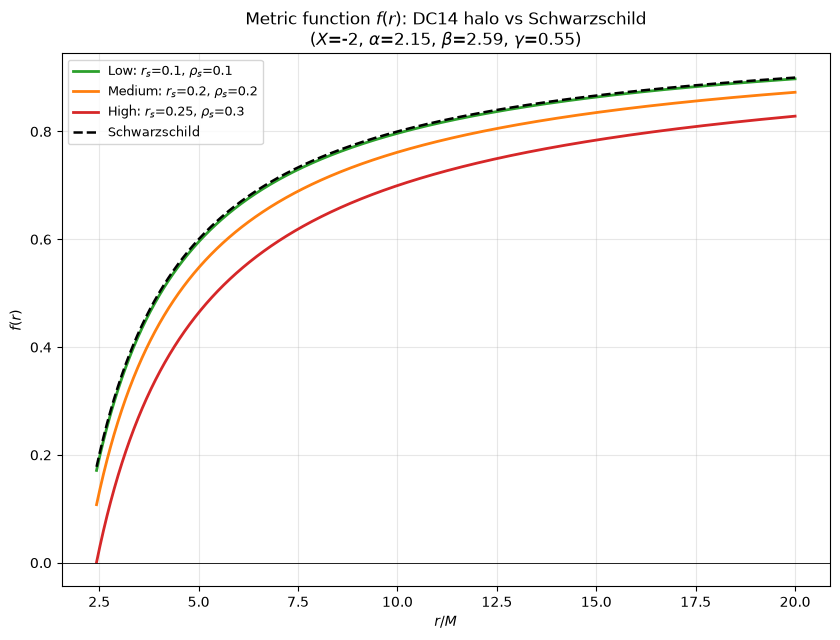}
    \caption{Metric function $f(r)$ for a Schwarzschild black hole surrounded by a DC14 dark-matter halo for a stellar-to-halo mass ratio of $X=-2$.}
    \label{fig:fr_dc14_Xm2}
\end{figure}

\subsection{Effective Potential}

The enclosed dark-matter mass associated with the DC14 profile is
obtained from
\begin{equation}
M_{\rm DC14}(r)
=
4\pi
\int_0^r
\rho_{\rm DC14}(\tilde r)\,
\tilde r^2\,d\tilde r .
\label{eq:dc14_mass}
\end{equation}

The corresponding metric function is
\begin{equation}
f_{\rm DC14}(r)
=
1-\frac{2\left[M+M_{\rm DC14}(r)\right]}{r}.
\label{eq:dc14_metric}
\end{equation}

For axial gravitational perturbations, the effective potential is
\begin{equation}
\begin{split}
V_{\rm DC14}^{(-)}(r)
=
f_{\rm DC14}(r)
\Bigg[
&\frac{\ell(\ell+1)}{r^2}
-\frac{6\left[M+M_{\rm DC14}(r)\right]}{r^3}
\\
&+\frac{2M_{\rm DC14}'(r)}{r^2}
-\frac{2M_{\rm DC14}''(r)}{r}
\Bigg].
\end{split}
\label{eq:dc14_potential}
\end{equation}

Here, the DC14 parameter $X$ determines the profile parameters
$(\alpha,\beta,\gamma)$ through
Eqs.~\eqref{eq:dc14_alpha}--\eqref{eq:dc14_gamma}. Thus,
$(\alpha,\beta,\gamma)$ are not varied independently. Instead, we
vary $X$, together with the scale radius $r_s$ and density
normalization $\rho_s$, to investigate their effects on the effective
potential and, subsequently, on the quasinormal-mode spectrum.

\begin{figure}[H]
    \centering
    \includegraphics[width=0.85\columnwidth]{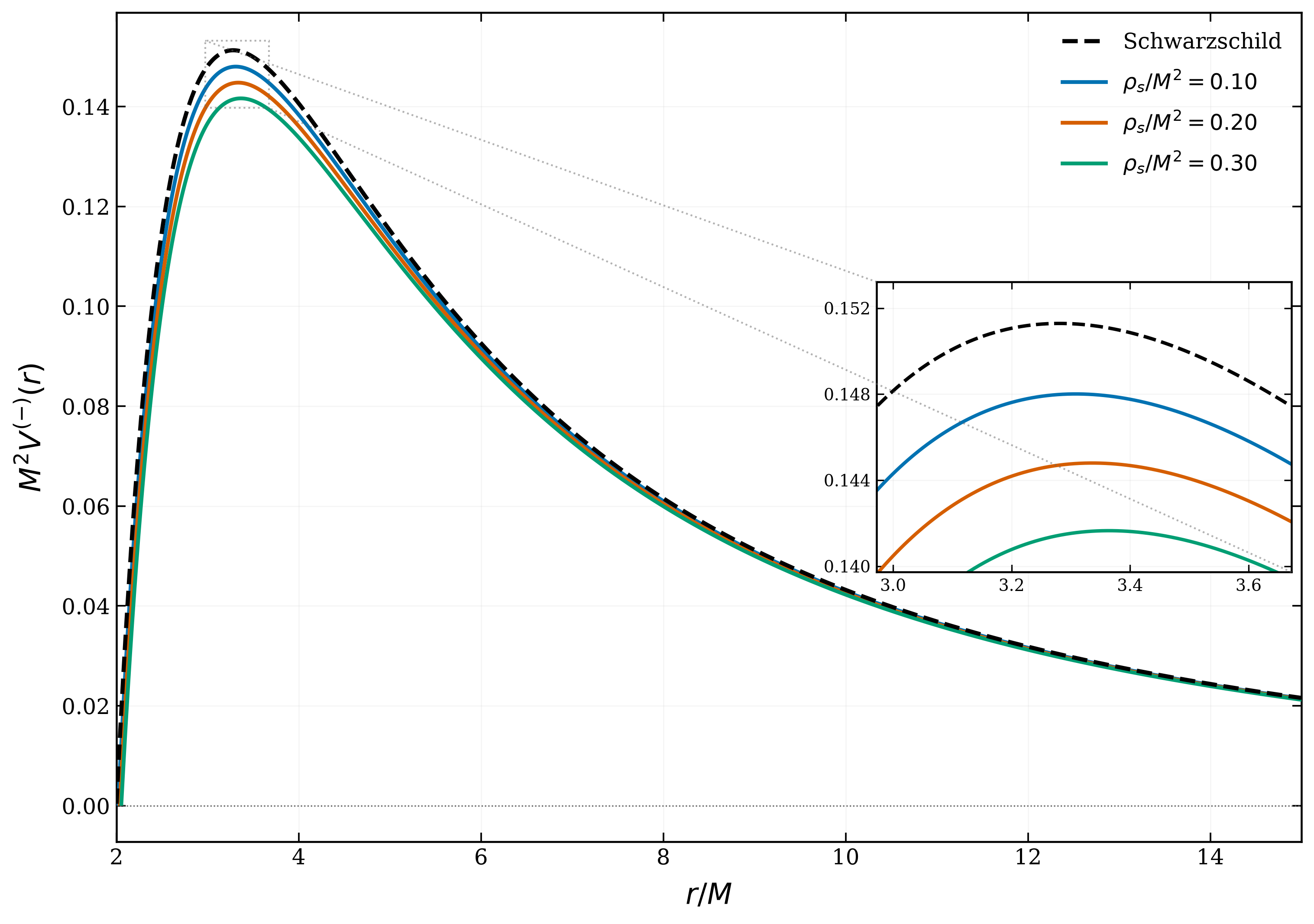}
    \caption{Dependence of the axial gravitational effective potential $V_{\rm DC14}^{(-)}(r)$ on the halo density normalization $\rho_s$ for $\ell=2$ and $X=-2.6$, shown for variation with $\rho_s$ at a fixed scale radius of $r_s/M=0.1$.}
    \label{fig:dc14_rhos}
\end{figure}

\begin{figure}[H]
    \centering
    \includegraphics[width=0.85\columnwidth]{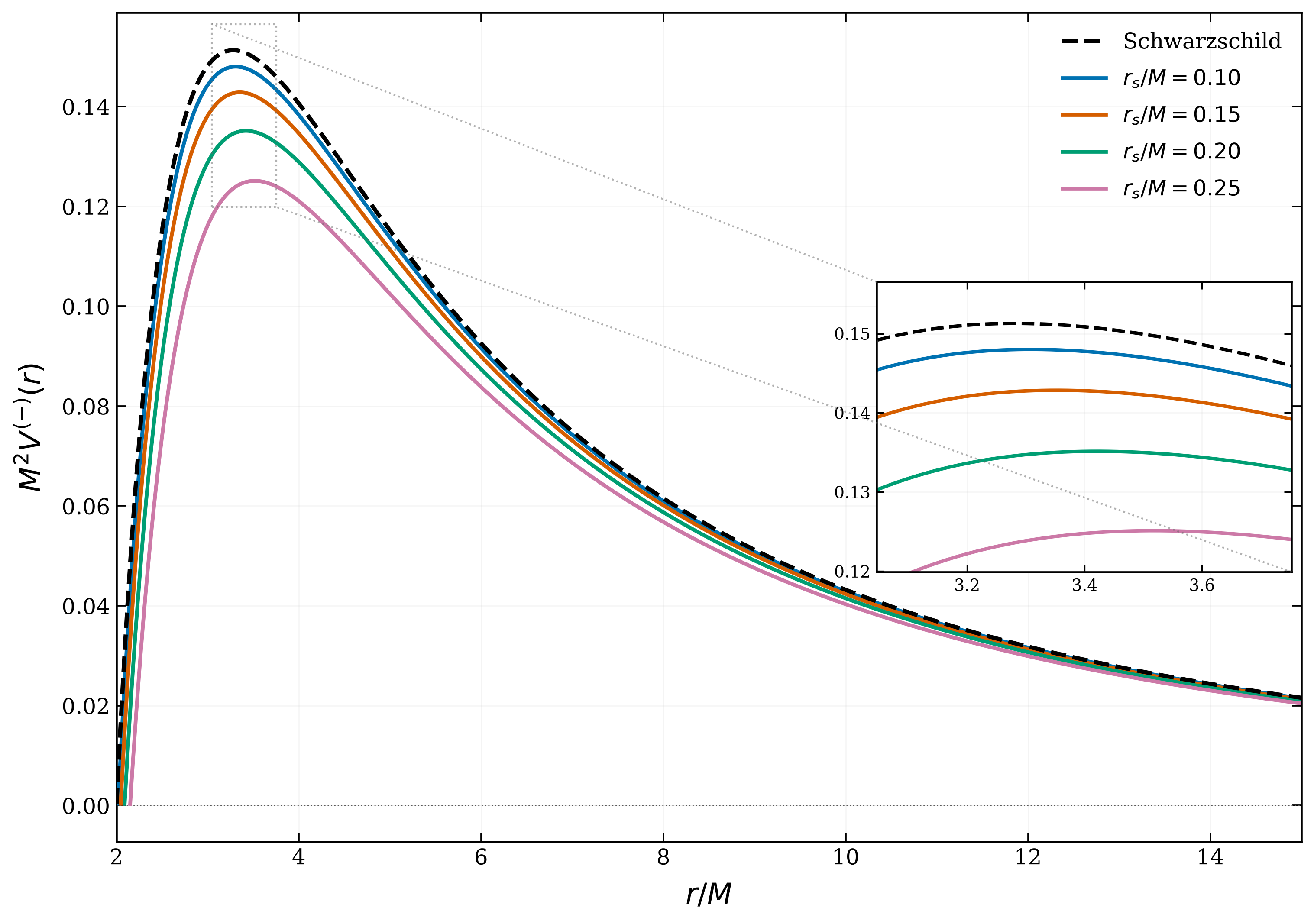}
    \caption{Dependence of the axial gravitational effective potential $V_{\rm DC14}^{(-)}(r)$ on the scale radius $r_s$ for $\ell=2$ and $X=-2.6$, shown for variation with $r_s$ at a fixed density normalization of $\rho_s/M^2=0.1$.}
    \label{fig:dc14_rs}
\end{figure}

Figs.~\ref{fig:dc14_rhos} and \ref{fig:dc14_rs} illustrate the dependence of the
potential barrier on the characteristic properties of the DC14 halo. Increasing either $\rho_s$ or $r_s$ systematically modifies the barrier relative to the vacuum Schwarzschild case, changing both its radial location and height. These changes provide the basis for studying the corresponding environmental corrections to the quasinormal-mode spectrum.

\begin{figure}[H]
    \centering
    \includegraphics[width=1\linewidth]{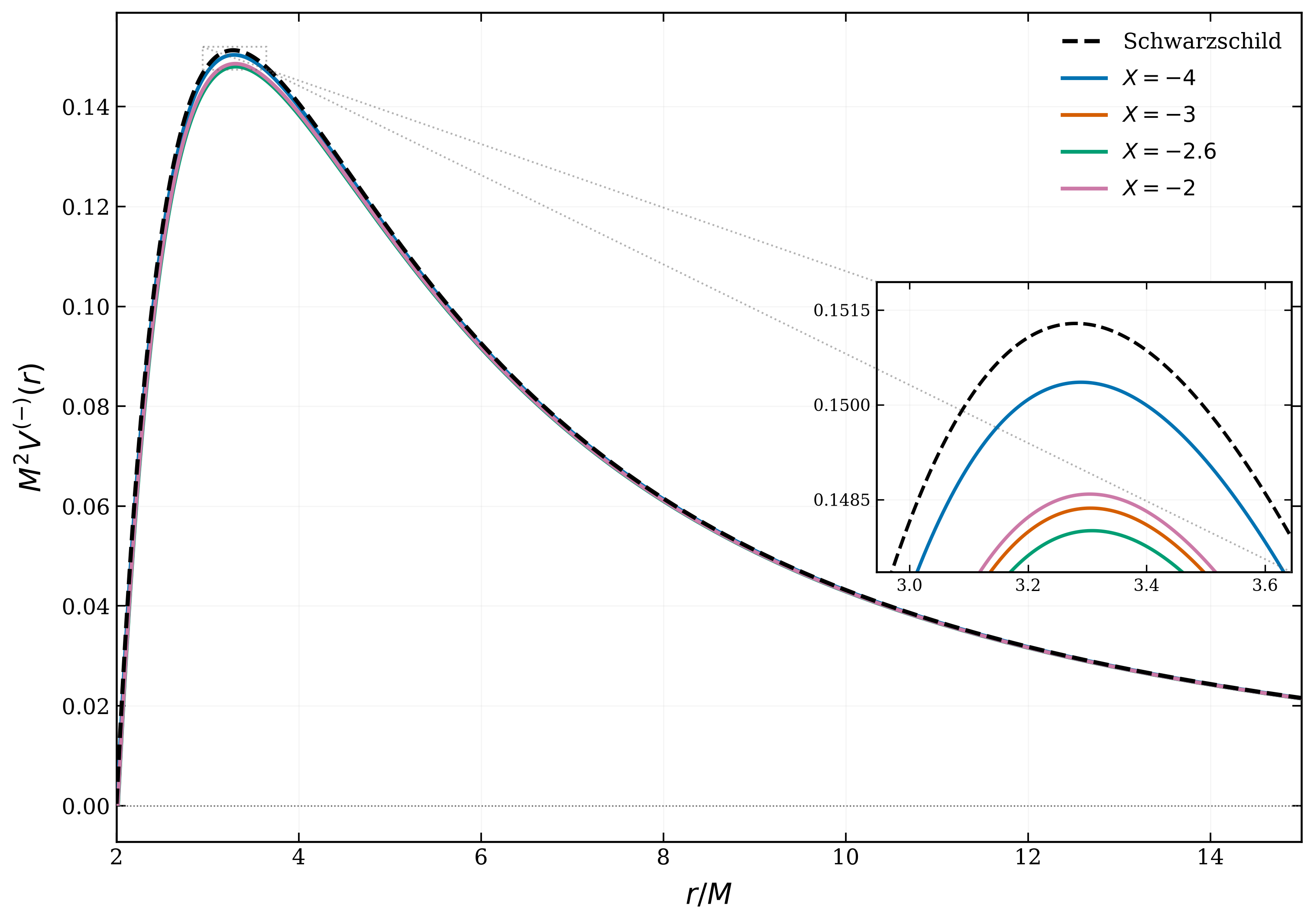}
    \caption{Dependence of the axial gravitational effective potential $V_{\rm DC14}^{(-)}(r)$ on the DC14 parameter $X$ for $\ell=2$. This panel corresponds to a halo density normalization of $\rho_s/M^2=0.1$ and a scale radius of $r_s/M=0.1$. Since $X$ determines the shape parameters $(\alpha,\beta,\gamma)$ through the DC14 relations, varying $X$ changes the halo profile and modifies the gravitational potential barrier. The Schwarzschild potential is included as a reference.}
    \label{fig:dc14_X_01}
\end{figure}
\begin{figure}[H]
    \centering
    \includegraphics[width=1\linewidth]{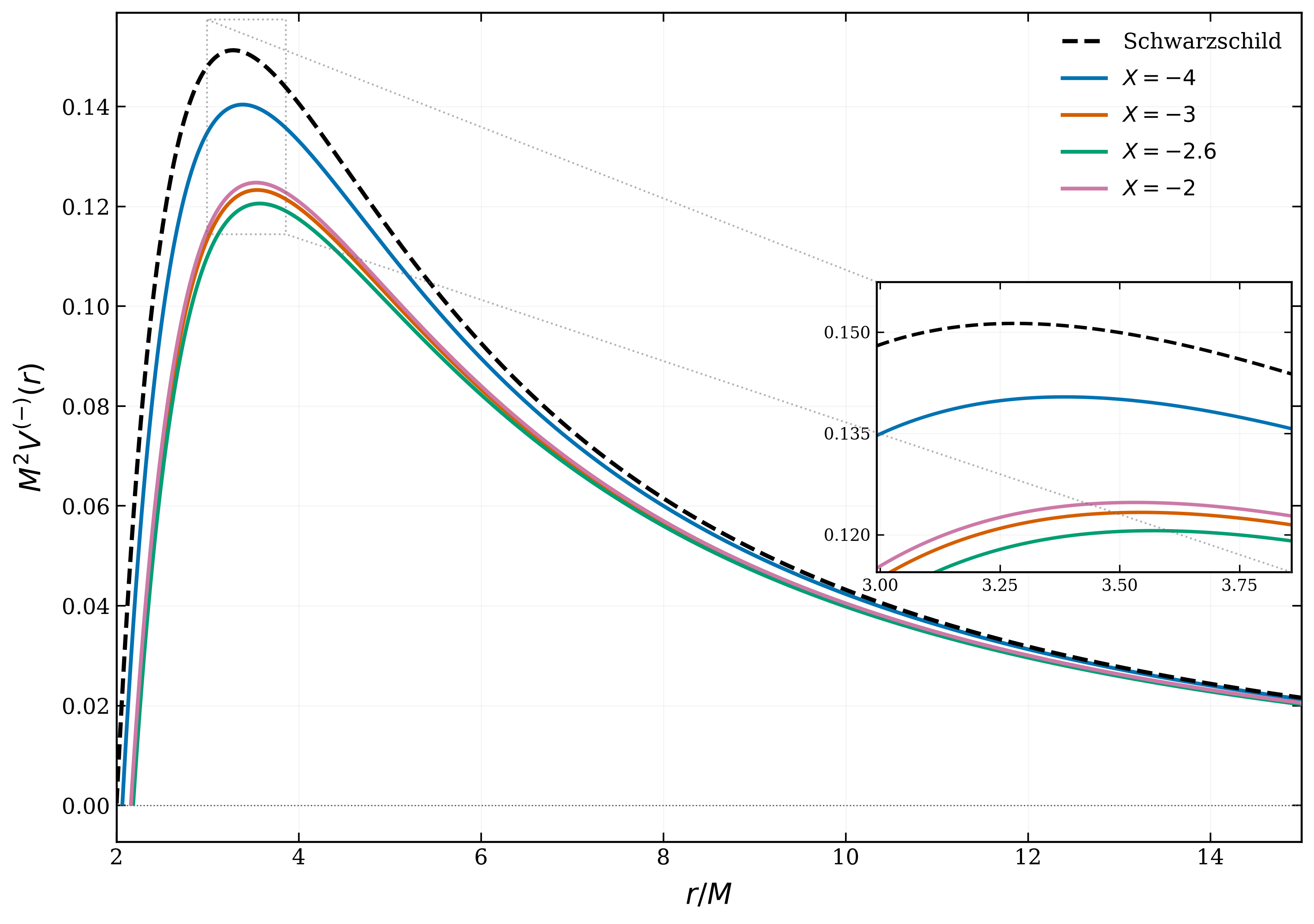}
    \caption{Dependence of the axial gravitational effective potential $V_{\rm DC14}^{(-)}(r)$ on the DC14 parameter $X$ for $\ell=2$. This panel corresponds to a halo density normalization of $\rho_s/M^2=0.2$ and a scale radius of $r_s/M=0.2$.}
    \label{fig:dc14_X_02}
\end{figure}

\begin{figure}[H]
    \centering
    \includegraphics[width=\linewidth]{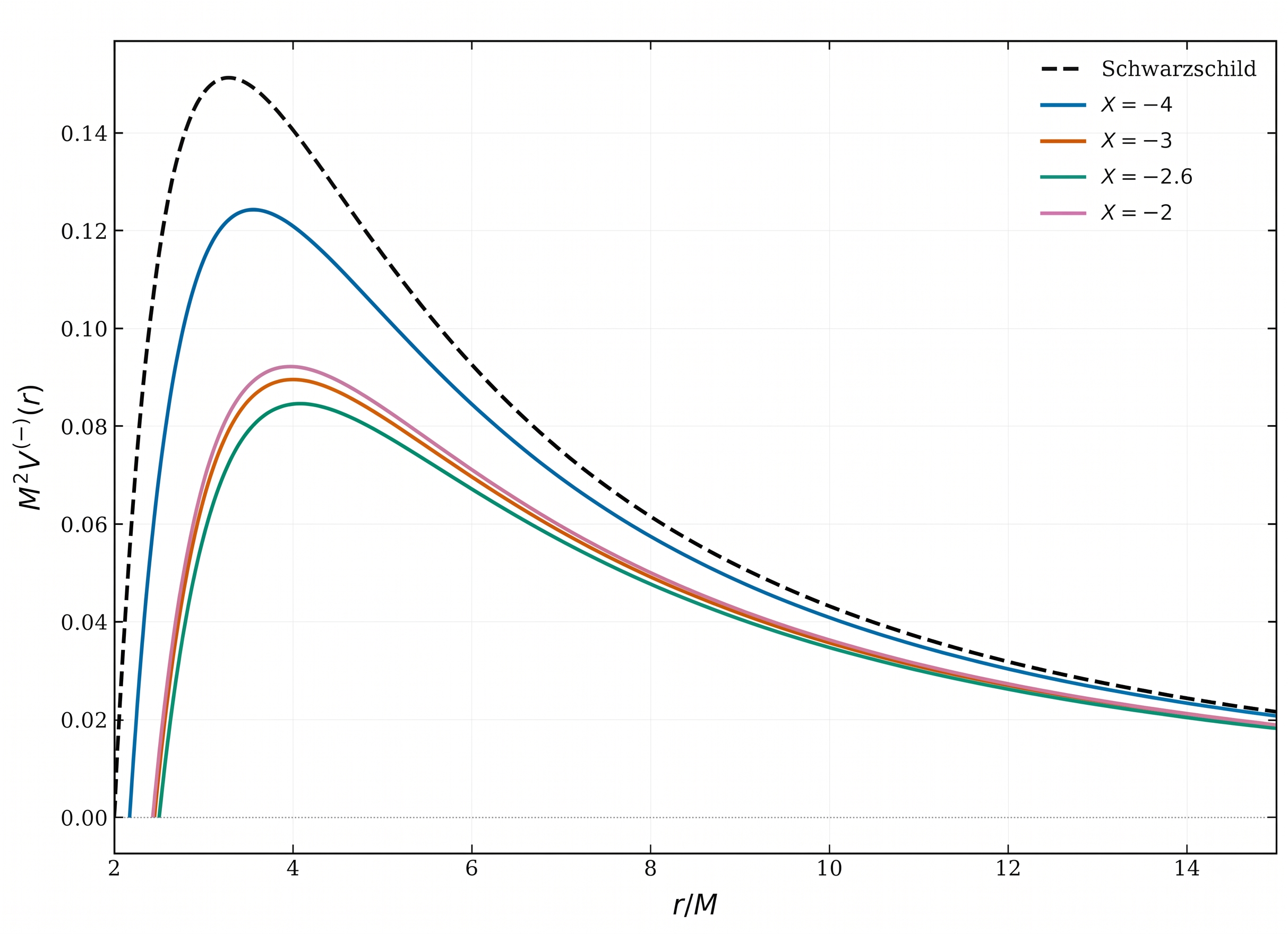}
    \caption{Dependence of the axial gravitational effective potential $V_{\rm DC14}^{(-)}(r)$ on the DC14 parameter $X$ for $\ell=2$. This panel corresponds to a halo density normalization of $\rho_s/M^2=0.3$ and a scale radius of $r_s/M=0.25$.}
    \label{fig:dc14_X_03}
\end{figure}
\vspace{-5pt}
\begin{figure}[H]
    \centering
    \includegraphics[width=\linewidth]
    {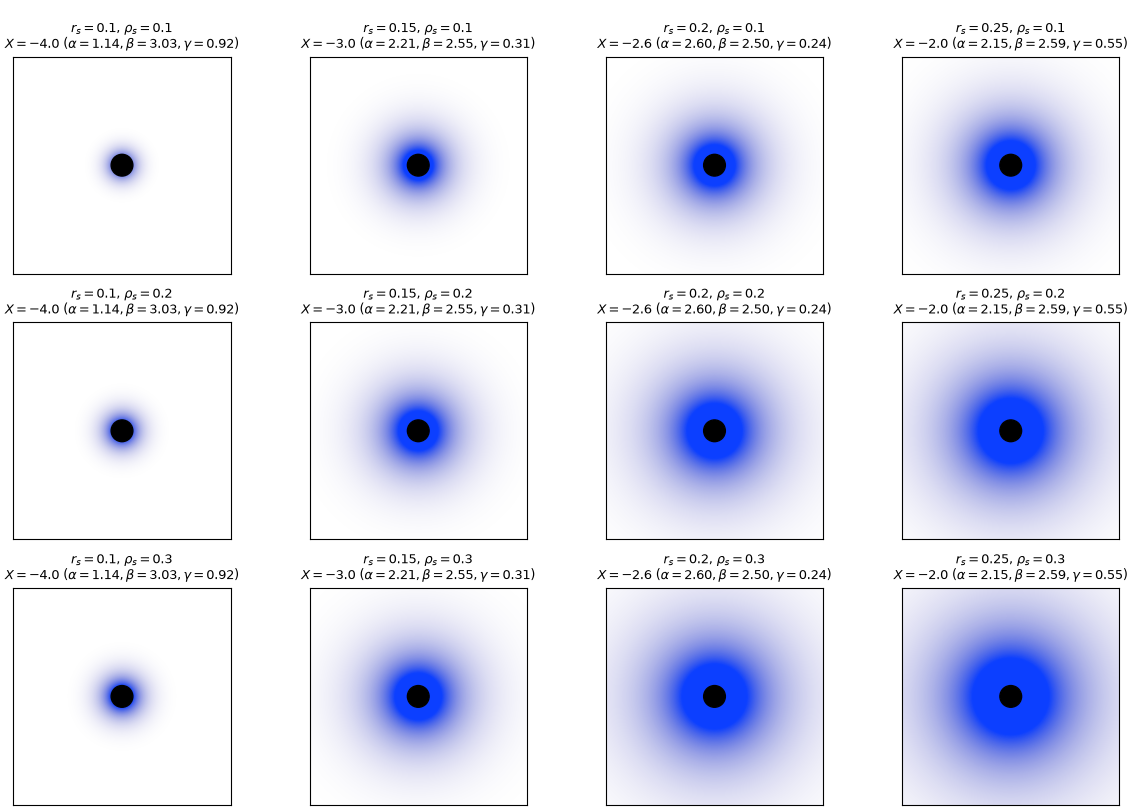}

    \caption{
    Schematic visualization of a Schwarzschild black hole surrounded
    by a DC14 dark-matter halo for the representative parameter
    combinations considered in this work. The four columns correspond
    to four selected values of the stellar-to-halo mass ratio,
    $X=-4.0,\,-3.0,\,-2.6,\,-2.0$, with corresponding fixed scale
    radii $r_s=0.10,\ 0.15,\ 0.20,\ 0.25$, respectively. For each
    column, the density normalization is varied over
    $\rho_s=0.10,\ 0.20,\ 0.30$ from top to bottom. The DC14 shape
    parameters $(\alpha,\beta,\gamma)$ are determined by $X$, while
    $r_s$ and $\rho_s$ specify the characteristic scale and density
    normalization of the halo. The black central region represents
    the black hole, while the surrounding blue distribution
    schematically represents the dark-matter halo. The visualization
    is schematic and is not intended to represent a physical
    two-dimensional projection or an absolute density scale.
    }
    \label{fig:dc14_halo_parameter_scan}
\end{figure}
The dependence on the DC14 parameter $X$ is shown in
Figs.~\ref{fig:dc14_X_01}--\ref{fig:dc14_X_03}. Since varying $X$ modifies the halo structure and hence the effective potential, it leads to corresponding shifts in the quasinormal-mode frequencies.


\section{Results}
\label{sec:Results}

\subsection{QNMs for Dehnen dark-matter halo profile}

\begin{table*}[htbp]
\centering

\caption{
Comparison between the fundamental $\ell=2$, $s=2$ QNM frequencies
obtained from time-domain evolution with Prony extraction and the
sixth-order WKB approximation for the Dehnen dark-matter halo.
The relative differences in the real and imaginary parts are denoted
by $\delta_R$ and $\delta_I$, respectively.
}

\label{tab:prony_wkb_dehnen}

\renewcommand{\arraystretch}{1.2}

\begin{tabular}{c c @{\hspace{2em}} c c @{\hspace{2em}} c c}
\toprule
$r_s/M$ & $\rho_s/M^2$
& $\omega_{\rm Prony}$
& $\omega_{\rm WKB}^{(6)}$
& $\delta_R$
& $\delta_I$
\\
\midrule


0.10 & 0.10 &
$0.372816-0.088686i$ &
$0.372701-0.088670i$ &
$0.03085\%$ &
$0.01849\%$
\\

0.15 & 0.10 &
$0.370678-0.088156i$ &
$0.370565-0.088151i$ &
$0.03048\%$ &
$0.00613\%$
\\

0.20 & 0.10 &
$0.366629-0.087169i$ &
$0.366522-0.087162i$ &
$0.02946\%$ &
$0.00860\%$
\\

0.25 & 0.10 &
$0.360203-0.085602i$ &
$0.360107-0.085585i$ &
$0.02665\%$ &
$0.01998\%$
\\

\cmidrule{1-6}


0.10 & 0.20 &
$0.371904-0.088453i$ &
$0.371786-0.088449i$ &
$0.03173\%$ &
$0.00430\%$
\\

0.15 & 0.20 &
$0.367669-0.087435i$ &
$0.367559-0.087422i$ &
$0.02992\%$ &
$0.01464\%$
\\

0.20 & 0.20 &
$0.359778-0.085507i$ &
$0.359678-0.085497i$ &
$0.02779\%$ &
$0.01169\%$
\\

0.25 & 0.20 &
$0.347589-0.082527i$ &
$0.347501-0.082512i$ &
$0.02532\%$ &
$0.01805\%$
\\

\cmidrule{1-6}


0.10 & 0.30 &
$0.370990-0.088243i$ &
$0.370877-0.088230i$ &
$0.03046\%$ &
$0.01473\%$
\\

0.15 & 0.30 &
$0.364707-0.086719i$ &
$0.364599-0.086706i$ &
$0.02961\%$ &
$0.01534\%$
\\

0.20 & 0.30 &
$0.353168-0.083907i$ &
$0.353077-0.083894i$ &
$0.02577\%$ &
$0.01525\%$
\\

0.25 & 0.30 &
$0.335801-0.079667i$ &
$0.335717-0.079649i$ &
$0.02501\%$ &
$0.02259\%$
\\

\bottomrule
\end{tabular}

\end{table*}
We first study the fundamental $\ell=2$, $s=2$ gravitational QNMs of a black hole surrounded by a Dehnen dark-matter halo. We vary the halo scale radius $r_s/M$ and density normalization $\rho_s/M^2$, obtaining the QNM frequencies from time-domain evolution followed by Prony extraction. The results are cross-checked using the sixth-order WKB approximation with the same effective potential.

\begin{figure}[H]
    \centering
    \includegraphics[width=0.95\columnwidth]{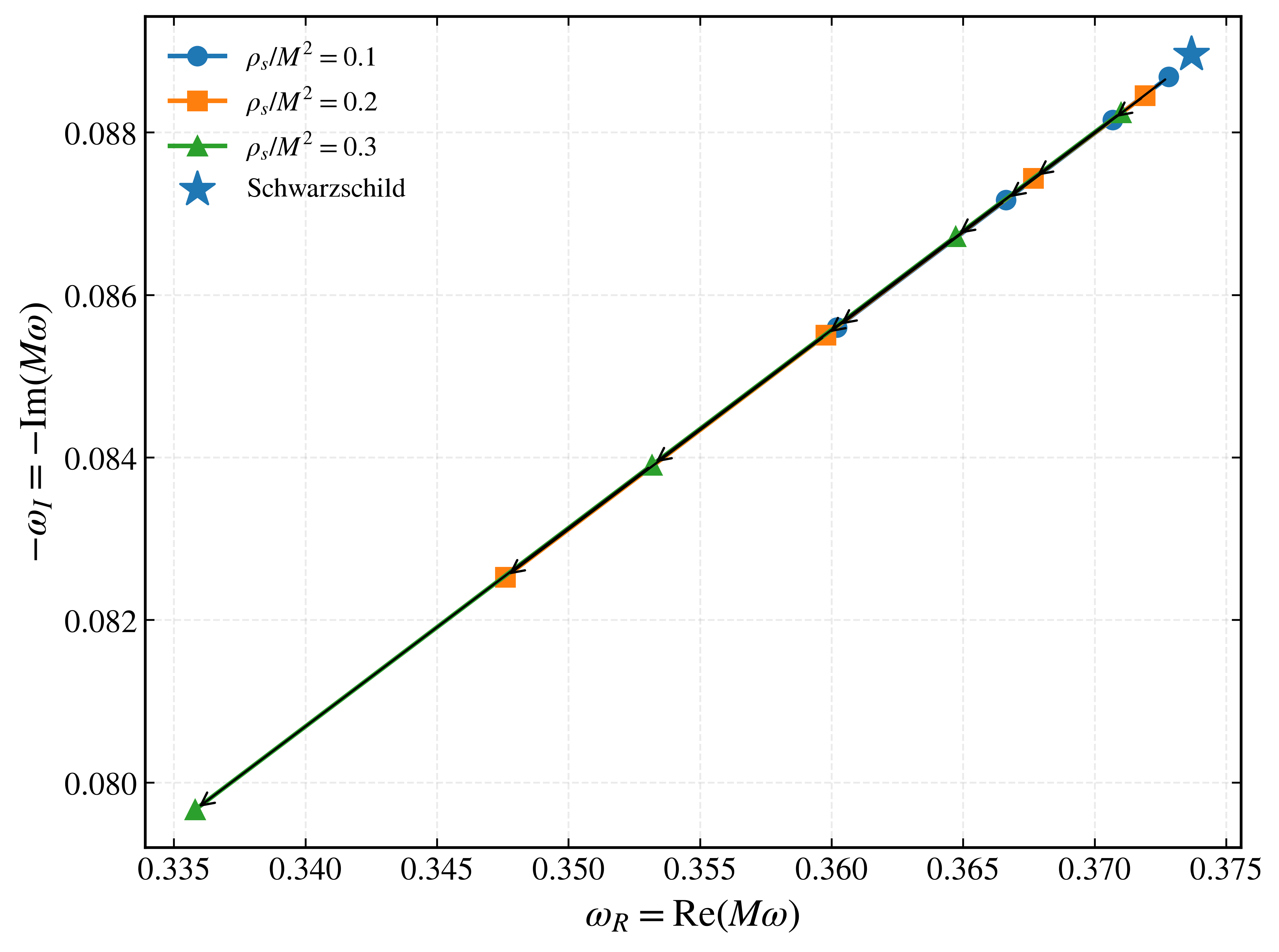}
    \caption{Fundamental $\ell=2$, $s=2$ gravitational QNM frequencies in the complex-frequency plane for the Dehnen dark-matter halo. The trajectories show the variation with $r_s/M$ for different $\rho_s/M^2$, with the Schwarzschild QNM shown for reference. 
    }
    \label{fig:dehnen_qnm_complex_plane}
\end{figure}

\begin{figure}[H]
    \centering
    \includegraphics[width=0.95\columnwidth]{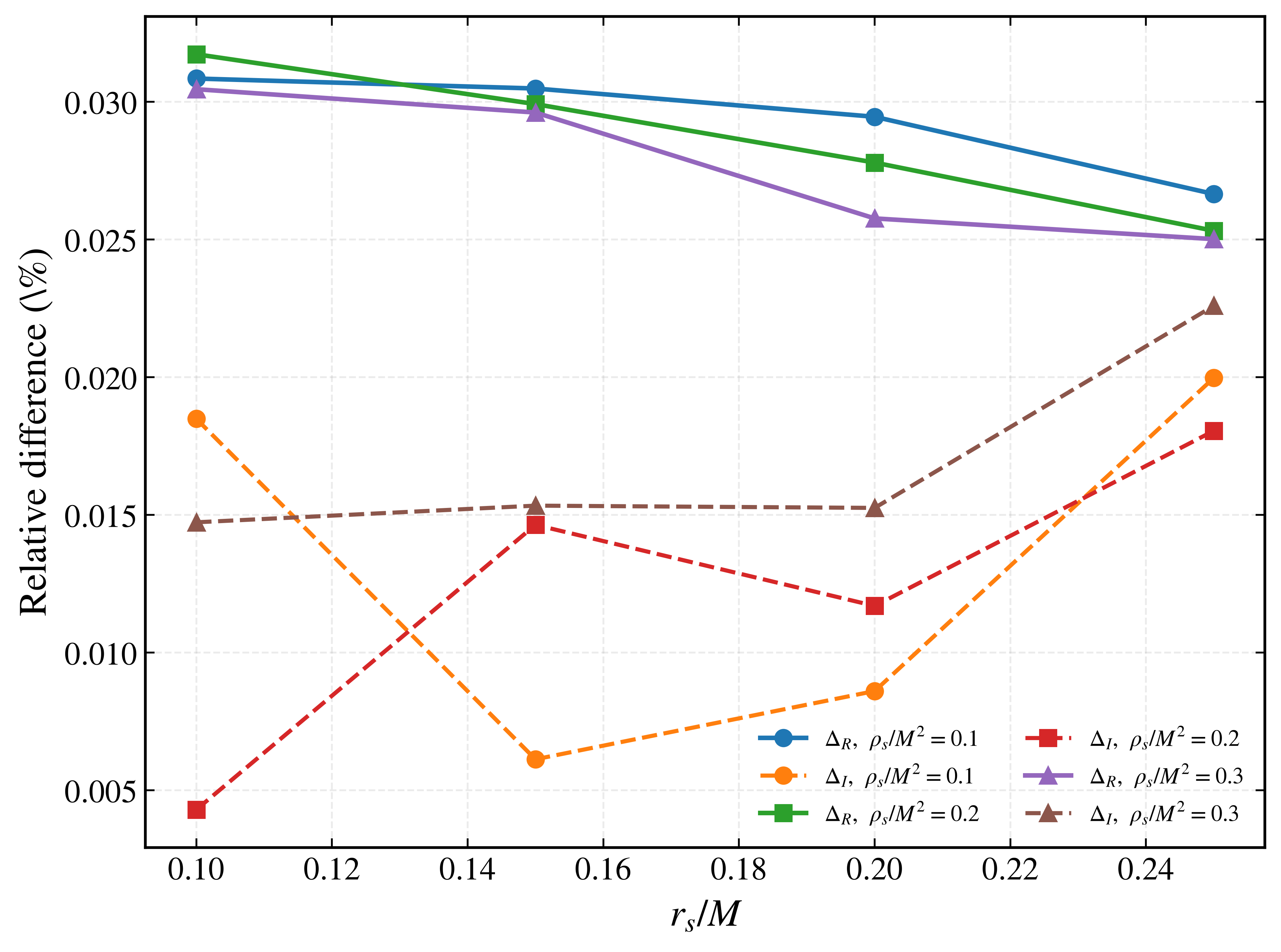}
    \caption{
    Relative differences $\delta_R$ and $\delta_I$ between the
    fundamental $\ell=2$, $s=2$ QNM frequencies obtained from
    time-domain evolution with Prony extraction and the sixth-order
    WKB approximation for the Dehnen dark-matter halo, shown as a
    function of $r_s/M$ for $\rho_s/M^2=0.1$, $0.2$, and $0.3$.
    }
    \label{fig:dehnen_prony_wkb_error}
\end{figure}

Fig.~\ref{fig:dehnen_qnm_complex_plane} shows a clear and systematic displacement of the QNM frequencies away from the Schwarzschild value. Increasing the halo density normalization $\rho_s/M^2$ and scale radius $r_s/M$ generally shifts the fundamental mode toward smaller $\omega_R$ and smaller damping rate $-\omega_I$. The displacement becomes more pronounced with increasing $r_s/M$, while the dependence on $\rho_s/M^2$ is comparatively weaker over the parameter range considered. Thus, the Dehnen halo parameters produce systematic shifts in the QNM spectrum, with the higher values of $r_s/M$ producing larger deviations from the vacuum Schwarzschild case. The corresponding Prony-extracted and sixth-order WKB frequencies are summarized in Table~\ref{tab:prony_wkb_dehnen}.

\subsection{QNMs for DC14 dark-matter halo profile}


We investigate the effect of a DC14 dark-matter halo on the
fundamental $\ell=2$, $s=2$ QNM. We vary $r_s/M$, $\rho_s/M^2$,
and the DC14 parameter $X$, and compare the resulting frequencies with the Schwarzschild value. The Prony-extracted frequencies are also cross-checked using the sixth-order WKB approximation.

\begin{figure}[H]
    \centering
    \includegraphics[width=\columnwidth]{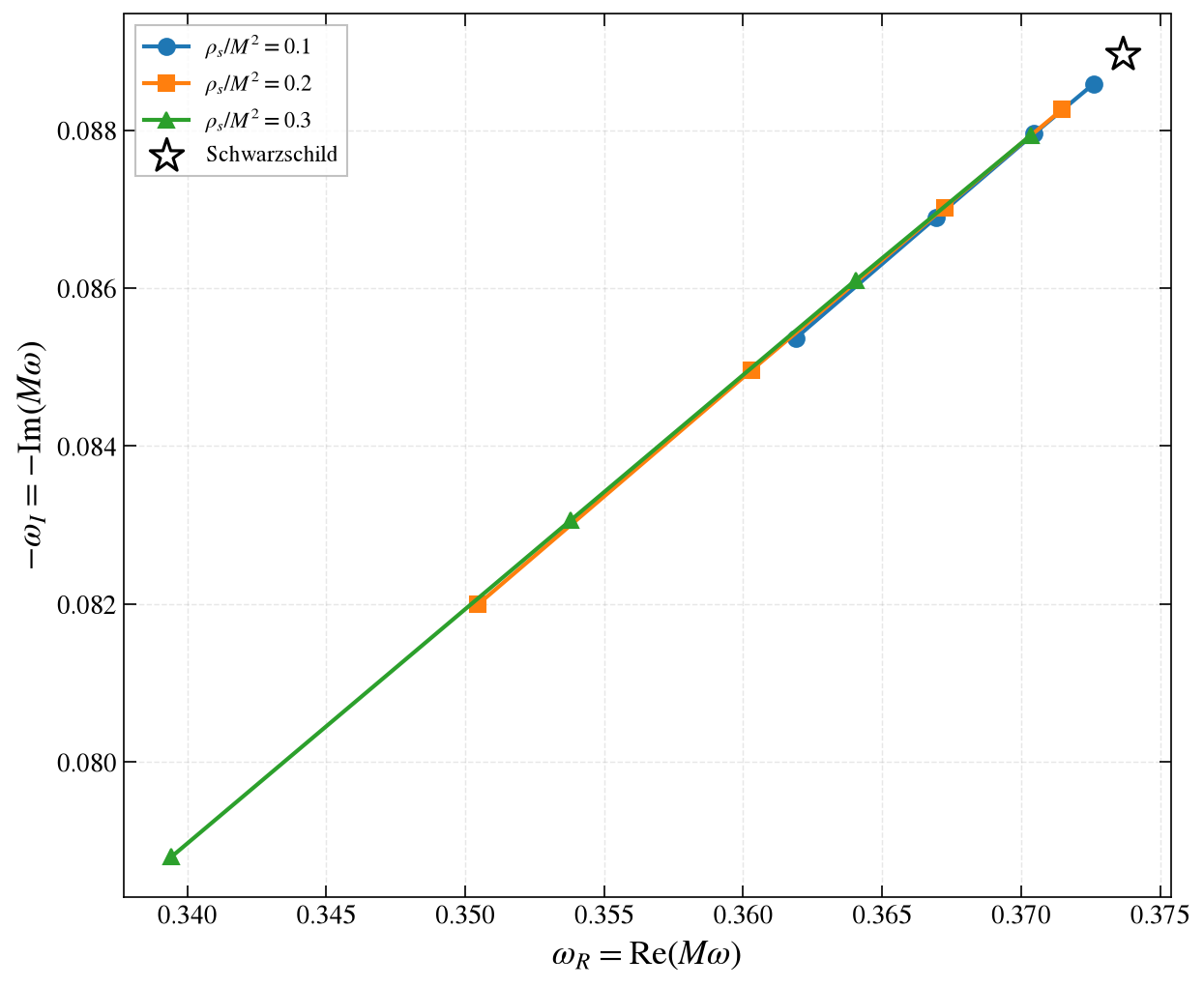}
    \caption{
    Fundamental $\ell=2$, $s=2$ gravitational QNM frequencies in the
    complex-frequency plane for the DC14 dark-matter halo at $X=-4$.
    The trajectories show the variation of the QNM frequency with the
    halo scale radius $r_s/M$ for the three density normalizations
    $\rho_s/M^2=0.1$, $0.2$, and $0.3$. The Schwarzschild QNM
    frequency is indicated by the star.
    }
    \label{fig:dc14_omega_complex_Xm4}
\end{figure}

\begin{figure}[H]
    \centering
    \includegraphics[width=0.8\columnwidth]{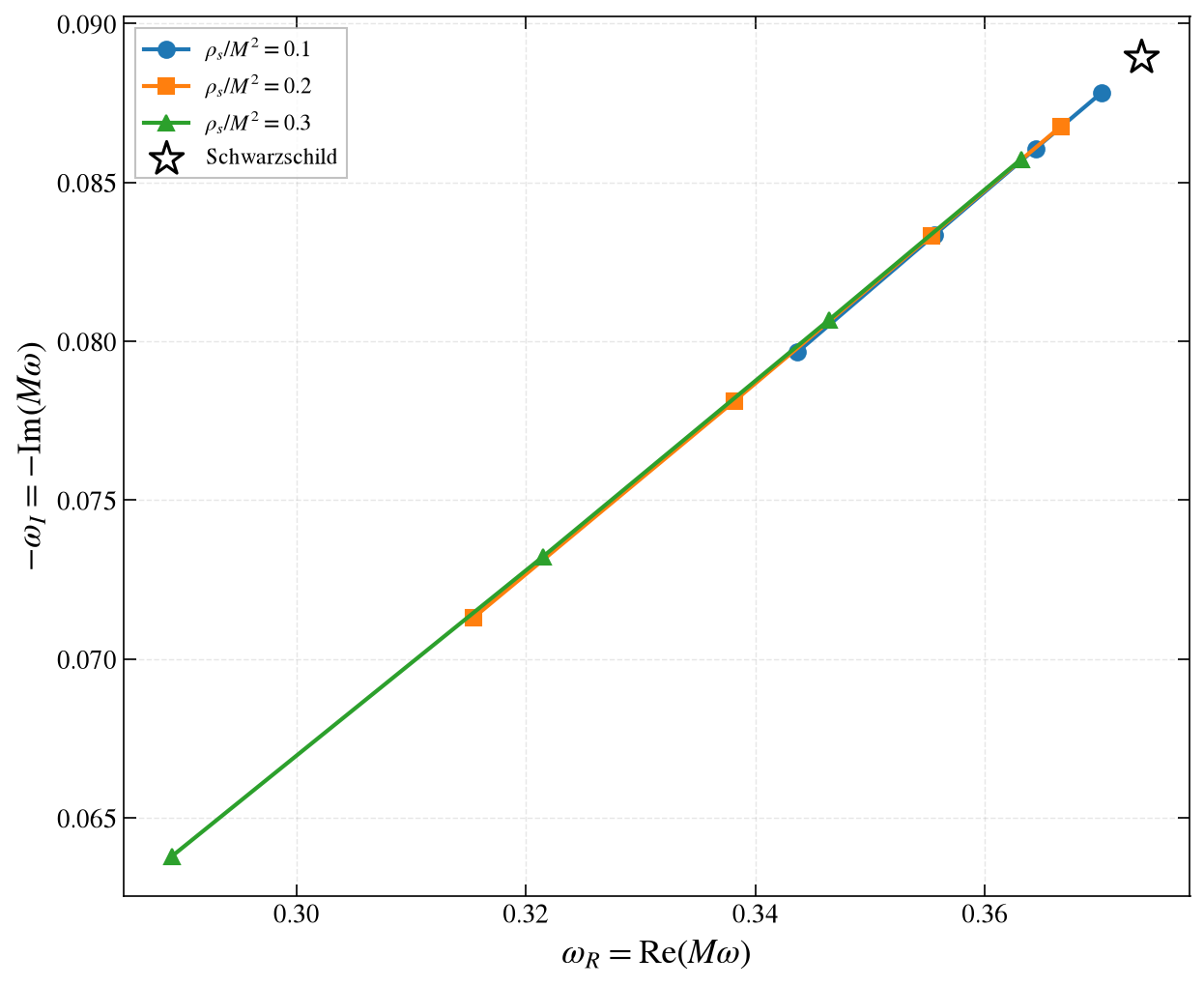}
    \caption{
    Same as Fig.~\ref{fig:dc14_omega_complex_Xm4}, for the DC14
    dark-matter halo at $X=-3$.
    }
    \label{fig:dc14_omega_complex_Xm3}
\end{figure}

\begin{figure}[H]
    \centering
    \includegraphics[width=0.8\columnwidth]{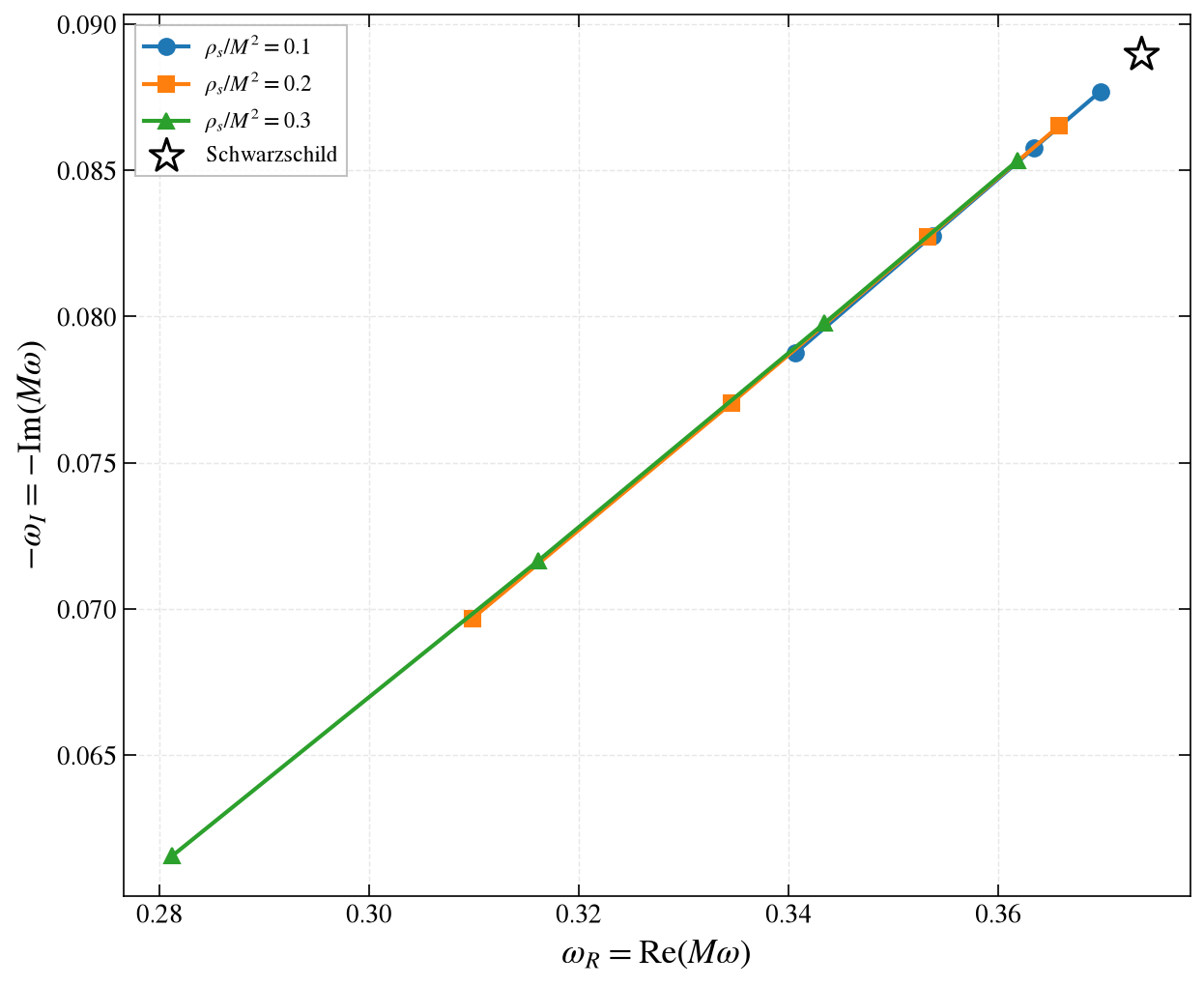}
    \caption{
    Same as Fig.~\ref{fig:dc14_omega_complex_Xm4}, for the DC14
    dark-matter halo at $X=-2.6$.
    }
    \label{fig:dc14_omega_complex_Xm2p6}
\end{figure}

\begin{figure}[H]
    \centering
    \includegraphics[width=0.85\columnwidth]{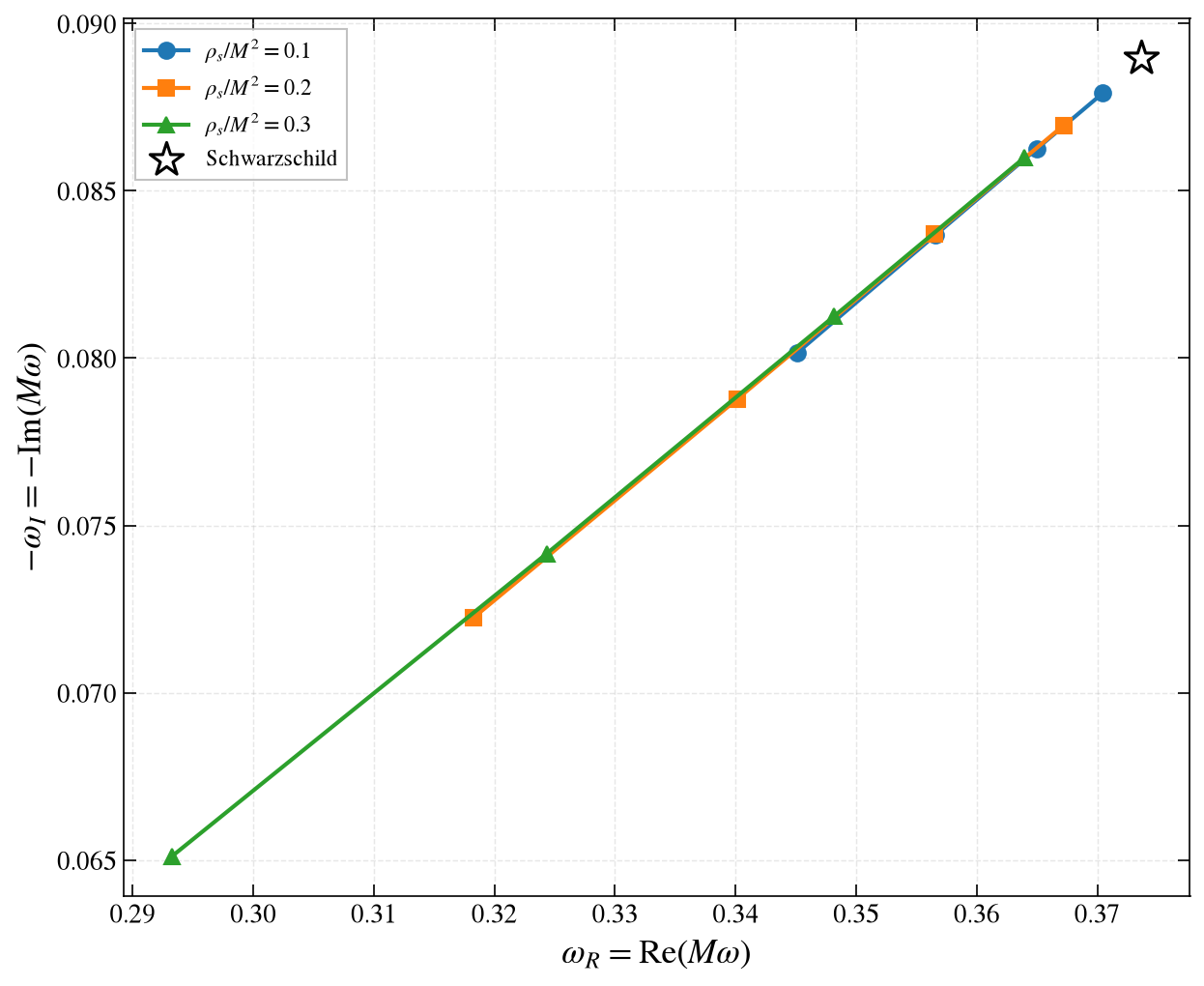}
    \caption{
    Same as Fig.~\ref{fig:dc14_omega_complex_Xm4}, for the DC14
    dark-matter halo at $X=-2$.
    }
    \label{fig:dc14_omega_complex_Xm2}
\end{figure}
\FloatBarrier

Figures~\ref{fig:dc14_omega_complex_Xm4}--\ref{fig:dc14_omega_complex_Xm2}
show a clear and systematic displacement of the QNM frequencies away
from the Schwarzschild value. Increasing the halo density normalization
$\rho_s/M^2$ and scale radius $r_s/M$ generally shifts the fundamental
mode toward smaller $\omega_R$ and smaller damping rate $-\omega_I$.
However, the dependence on the DC14 parameter $X$ is not monotonic.
Among the values considered, the largest displacement from the
Schwarzschild QNM is observed for $X=-2.6$, indicating that the
response of the QNM spectrum depends non-trivially on the inner
structure of the DC14 halo. Thus, both the halo parameters and the
value of $X$ influence the magnitude of the deviation from the
vacuum Schwarzschild spectrum.
\begin{table*}[htbp]
\centering

\caption{
Comparison between the fundamental $\ell=2$, $s=2$ QNM frequencies
obtained from time-domain evolution with Prony extraction and the
sixth-order WKB approximation for the DC14 halo at $X=-4$.
The relative differences in the real and imaginary parts are denoted
by $\delta_R$ and $\delta_I$, respectively.
}

\label{tab:prony_wkb_dc14_Xm4}

\renewcommand{\arraystretch}{1.2}

\begin{tabular}{c c @{\hspace{2em}} c c @{\hspace{2em}} c c}
\toprule
$r_s/M$ & $\rho_s/M^2$
& $\omega_{\rm Prony}$
& $\omega_{\rm WKB}^{(6)}$
& $\delta_R$
& $\delta_I$
\\
\midrule

0.10 & 0.10 & $0.372607-0.088585i$ & $0.372489-0.088569i$ & $0.032\%$ & $0.018\%$ \\
0.15 & 0.10 & $0.370475-0.087957i$ & $0.370365-0.087944i$ & $0.030\%$ & $0.015\%$ \\
0.20 & 0.10 & $0.366948-0.086895i$ & $0.366845-0.086887i$ & $0.028\%$ & $0.009\%$ \\
0.25 & 0.10 & $0.361898-0.085368i$ & $0.361807-0.085354i$ & $0.025\%$ & $0.017\%$ \\
\cmidrule{1-6}
0.10 & 0.20 & $0.371477-0.088263i$ & $0.371363-0.088249i$ & $0.031\%$ & $0.016\%$ \\
0.15 & 0.20 & $0.367248-0.087022i$ & $0.367144-0.087010i$ & $0.028\%$ & $0.013\%$ \\
0.20 & 0.20 & $0.360293-0.084957i$ & $0.360210-0.084941i$ & $0.023\%$ & $0.019\%$ \\
0.25 & 0.20 & $0.350468-0.081997i$ & $0.350392-0.081988i$ & $0.022\%$ & $0.011\%$ \\
\cmidrule{1-6}
0.10 & 0.30 & $0.370352-0.087936i$ & $0.370241-0.087930i$ & $0.030\%$ & $0.006\%$ \\
0.15 & 0.30 & $0.364055-0.086103i$ & $0.363958-0.086090i$ & $0.027\%$ & $0.015\%$ \\
0.20 & 0.30 & $0.353791-0.083059i$ & $0.353712-0.083051i$ & $0.022\%$ & $0.010\%$ \\
0.25 & 0.30 & $0.339432-0.078799i$ & $0.339369-0.078787i$ & $0.019\%$ & $0.015\%$ \\

\bottomrule
\end{tabular}
\end{table*}

\begin{table*}[htbp]
\centering

\caption{
Comparison between the fundamental $\ell=2$, $s=2$ QNM frequencies
obtained from time-domain evolution with Prony extraction and the
sixth-order WKB approximation for the DC14 halo at $X=-3$.
The relative differences in the real and imaginary parts are denoted
by $\delta_R$ and $\delta_I$, respectively.
}

\label{tab:prony_wkb_dc14_Xm3}

\renewcommand{\arraystretch}{1.2}

\begin{tabular}{c c @{\hspace{2em}} c c @{\hspace{2em}} c c}
\toprule
$r_s/M$ & $\rho_s/M^2$
& $\omega_{\rm Prony}$
& $\omega_{\rm WKB}^{(6)}$
& $\delta_R$
& $\delta_I$
\\
\midrule

0.10 & 0.10 & $0.370188-0.087828i$ & $0.370076-0.087822i$ & $0.0303\%$ & $0.0071\%$ \\
0.15 & 0.10 & $0.364436-0.086069i$ & $0.364338-0.086061i$ & $0.0269\%$ & $0.0091\%$ \\
0.20 & 0.10 & $0.355597-0.083342i$ & $0.355517-0.083335i$ & $0.0225\%$ & $0.0088\%$ \\
0.25 & 0.10 & $0.343654-0.079654i$ & $0.343589-0.079644i$ & $0.0189\%$ & $0.0126\%$ \\
\cmidrule{1-6}
0.10 & 0.20 & $0.366666-0.086773i$ & $0.366564-0.086766i$ & $0.0278\%$ & $0.0084\%$ \\
0.15 & 0.20 & $0.355333-0.083328i$ & $0.355253-0.083320i$ & $0.0225\%$ & $0.0096\%$ \\
0.20 & 0.20 & $0.338182-0.078121i$ & $0.338121-0.078110i$ & $0.0180\%$ & $0.0136\%$ \\
0.25 & 0.20 & $0.315472-0.071310i$ & $0.315433-0.071298i$ & $0.0124\%$ & $0.0175\%$ \\
\cmidrule{1-6}
0.10 & 0.30 & $0.363179-0.085735i$ & $0.363081-0.085723i$ & $0.0270\%$ & $0.0143\%$ \\
0.15 & 0.30 & $0.346432-0.080676i$ & $0.346362-0.080665i$ & $0.0202\%$ & $0.0132\%$ \\
0.20 & 0.30 & $0.321476-0.073217i$ & $0.321432-0.073204i$ & $0.0137\%$ & $0.0176\%$ \\
0.25 & 0.30 & $0.289184-0.063801i$ & $0.289164-0.063794i$ & $0.0069\%$ & $0.0111\%$ \\

\bottomrule
\end{tabular}
\end{table*}

\begin{table*}[htbp]
\centering

\caption{
Comparison between the fundamental $\ell=2$, $s=2$ QNM frequencies
obtained from time-domain evolution with Prony extraction and the
sixth-order WKB approximation for the DC14 halo at $X=-2.6$.
The relative differences in the real and imaginary parts are denoted
by $\delta_R$ and $\delta_I$, respectively.
}

\label{tab:prony_wkb_dc14_Xm2p6}

\renewcommand{\arraystretch}{1.2}

\begin{tabular}{c c @{\hspace{2em}} c c @{\hspace{2em}} c c}
\toprule
$r_s/M$ & $\rho_s/M^2$
& $\omega_{\rm Prony}$
& $\omega_{\rm WKB}^{(6)}$
& $\delta_R$
& $\delta_I$
\\
\midrule

0.10 & 0.10 & $0.369751-0.087693i$ & $0.369643-0.087687i$ & $0.0292\%$ & $0.0073\%$ \\
0.15 & 0.10 & $0.363401-0.085748i$ & $0.363305-0.085741i$ & $0.0264\%$ & $0.0085\%$ \\
0.20 & 0.10 & $0.353704-0.082763i$ & $0.353627-0.082754i$ & $0.0218\%$ & $0.0104\%$ \\
0.25 & 0.10 & $0.340665-0.078750i$ & $0.340603-0.078740i$ & $0.0182\%$ & $0.0126\%$ \\
\cmidrule{1-6}
0.10 & 0.20 & $0.365804-0.086510i$ & $0.365703-0.086498i$ & $0.0276\%$ & $0.0136\%$ \\
0.15 & 0.20 & $0.353301-0.082705i$ & $0.353223-0.082697i$ & $0.0221\%$ & $0.0103\%$ \\
0.20 & 0.20 & $0.334526-0.077021i$ & $0.334470-0.077011i$ & $0.0167\%$ & $0.0132\%$ \\
0.25 & 0.20 & $0.309834-0.069662i$ & $0.309798-0.069648i$ & $0.0116\%$ & $0.0197\%$ \\
\cmidrule{1-6}
0.10 & 0.30 & $0.361891-0.085342i$ & $0.361799-0.085326i$ & $0.0254\%$ & $0.0191\%$ \\
0.15 & 0.30 & $0.343439-0.079767i$ & $0.343373-0.079756i$ & $0.0192\%$ & $0.0140\%$ \\
0.20 & 0.30 & $0.316189-0.071657i$ & $0.316149-0.071643i$ & $0.0127\%$ & $0.0194\%$ \\
0.25 & 0.30 & $0.281240-0.061550i$ & $0.281223-0.061544i$ & $0.0060\%$ & $0.0094\%$ \\

\bottomrule
\end{tabular}
\end{table*}

\begin{table*}[htbp]
\centering

\caption{
Comparison between the fundamental $\ell=2$, $s=2$ QNM frequencies
obtained from time-domain evolution with Prony extraction and the
sixth-order WKB approximation for the DC14 halo at $X=-2$.
The relative differences in the real and imaginary parts are denoted
by $\delta_R$ and $\delta_I$, respectively.
}

\label{tab:prony_wkb_dc14_Xm2}

\renewcommand{\arraystretch}{1.2}

\begin{tabular}{c c @{\hspace{2em}} c c @{\hspace{2em}} c c}
\toprule
$r_s/M$ & $\rho_s/M^2$
& $\omega_{\rm Prony}$
& $\omega_{\rm WKB}^{(6)}$
& $\delta_R$
& $\delta_I$
\\
\midrule

0.10 & 0.10 & $0.370455-0.087916i$ & $0.370344-0.087910i$ & $0.0300\%$ & $0.0068\%$ \\
0.15 & 0.10 & $0.365025-0.086266i$ & $0.364927-0.086259i$ & $0.0268\%$ & $0.0081\%$ \\
0.20 & 0.10 & $0.356588-0.083680i$ & $0.356506-0.083673i$ & $0.0230\%$ & $0.0084\%$ \\
0.25 & 0.10 & $0.345095-0.080152i$ & $0.345027-0.080142i$ & $0.0197\%$ & $0.0126\%$ \\
\cmidrule{1-6}
0.10 & 0.20 & $0.367201-0.086952i$ & $0.367097-0.086941i$ & $0.0283\%$ & $0.0132\%$ \\
0.15 & 0.20 & $0.356498-0.083714i$ & $0.356416-0.083707i$ & $0.0230\%$ & $0.0079\%$ \\
0.20 & 0.20 & $0.340129-0.078771i$ & $0.340064-0.078761i$ & $0.0191\%$ & $0.0124\%$ \\
0.25 & 0.20 & $0.318286-0.072248i$ & $0.318245-0.072235i$ & $0.0129\%$ & $0.0179\%$ \\
\cmidrule{1-6}
0.10 & 0.30 & $0.363974-0.085995i$ & $0.363876-0.085983i$ & $0.0269\%$ & $0.0144\%$ \\
0.15 & 0.30 & $0.348159-0.081245i$ & $0.348086-0.081234i$ & $0.0210\%$ & $0.0135\%$ \\
0.20 & 0.30 & $0.324337-0.074155i$ & $0.324290-0.074142i$ & $0.0145\%$ & $0.0177\%$ \\
0.25 & 0.30 & $0.293286-0.065119i$ & $0.293263-0.065112i$ & $0.0078\%$ & $0.0103\%$ \\

\bottomrule
\end{tabular}
\end{table*}

\begin{figure}[H]
    \centering
    \includegraphics[width=\columnwidth] {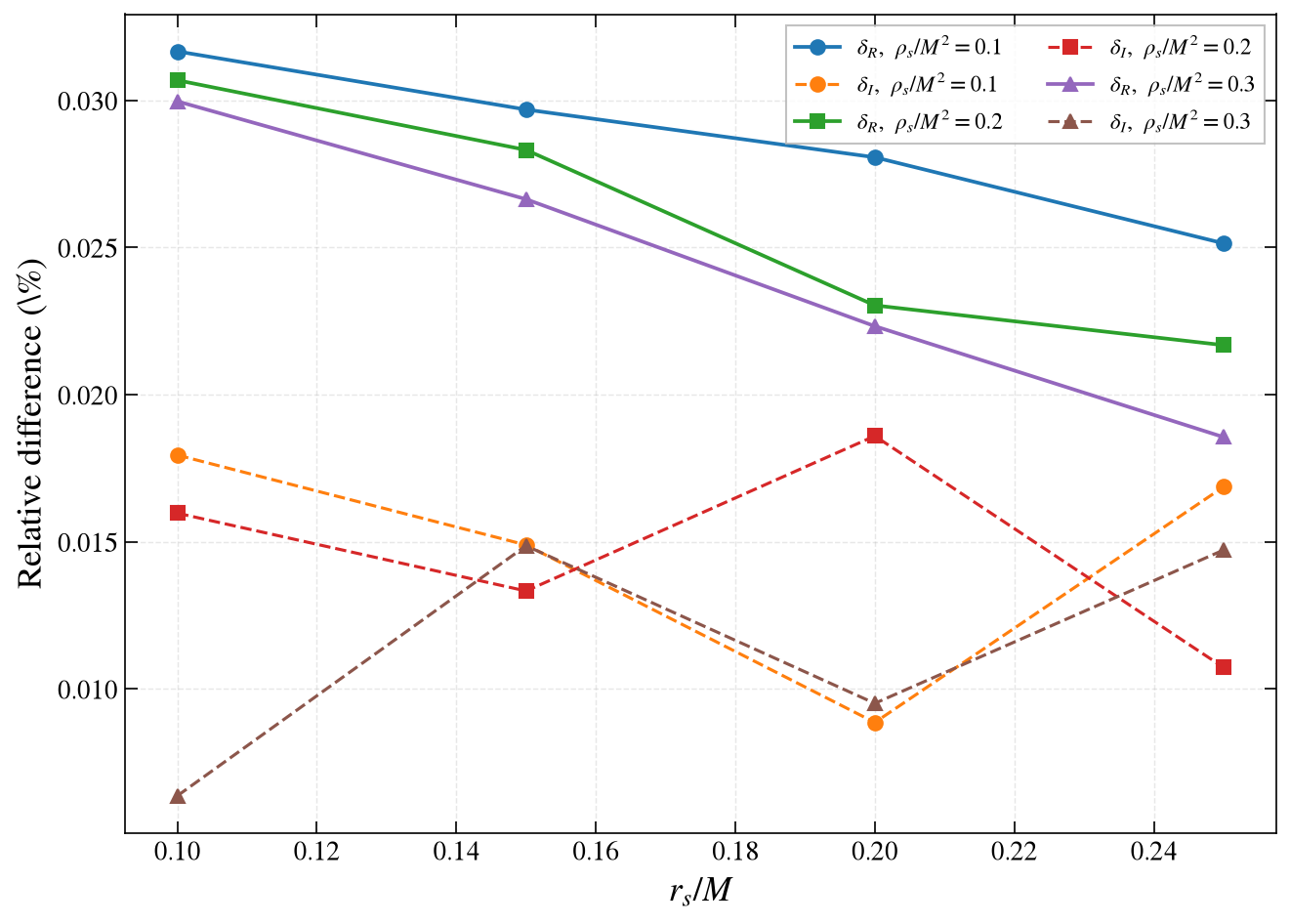}
    \caption{Relative differences between the Prony-extracted and sixth-order WKB QNM frequencies for the DC14 halo at $X=-4$ for $r_s/M=0.1, 0.15, 0.2, 0.25$ and $\rho_s/M^2=0.1$, $0.2$, and $0.3$. }
    \label{fig:dc14_error_xm4}
\end{figure}

\begin{figure}[H]
    \centering
    \includegraphics[width=\columnwidth]
    {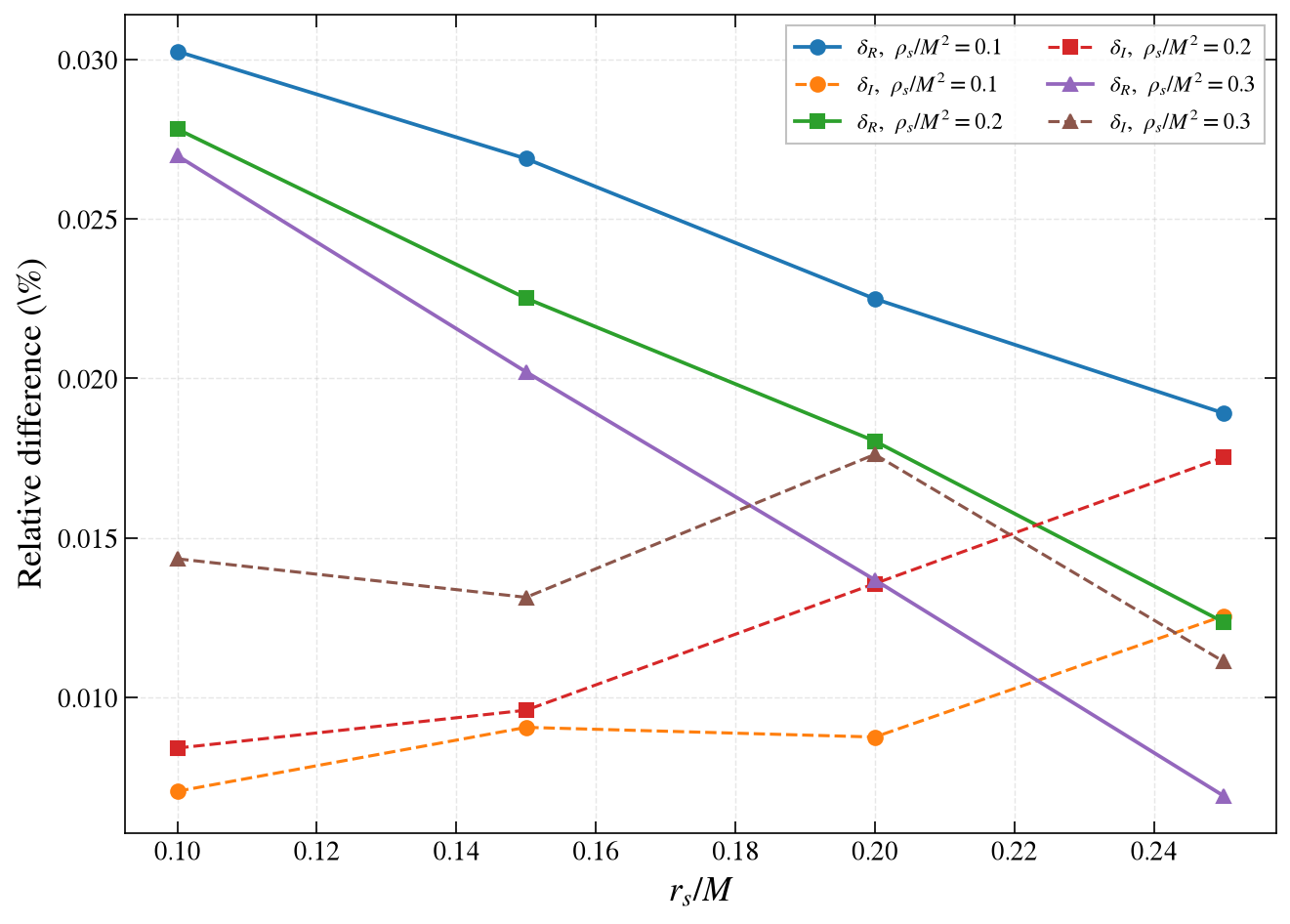}
    \caption{Relative differences between the Prony-extracted and sixth-order WKB QNM frequencies for the DC14 halo at $X=-3$ for  $r_s/M=0.1, 0.15, 0.2, 0.25$ and $\rho_s/M^2=0.1$, $0.2$, and $0.3$. }
    \label{fig:dc14_error_xm3}
\end{figure}
\begin{figure}[H]
    \centering
    \includegraphics[width=\columnwidth]
    {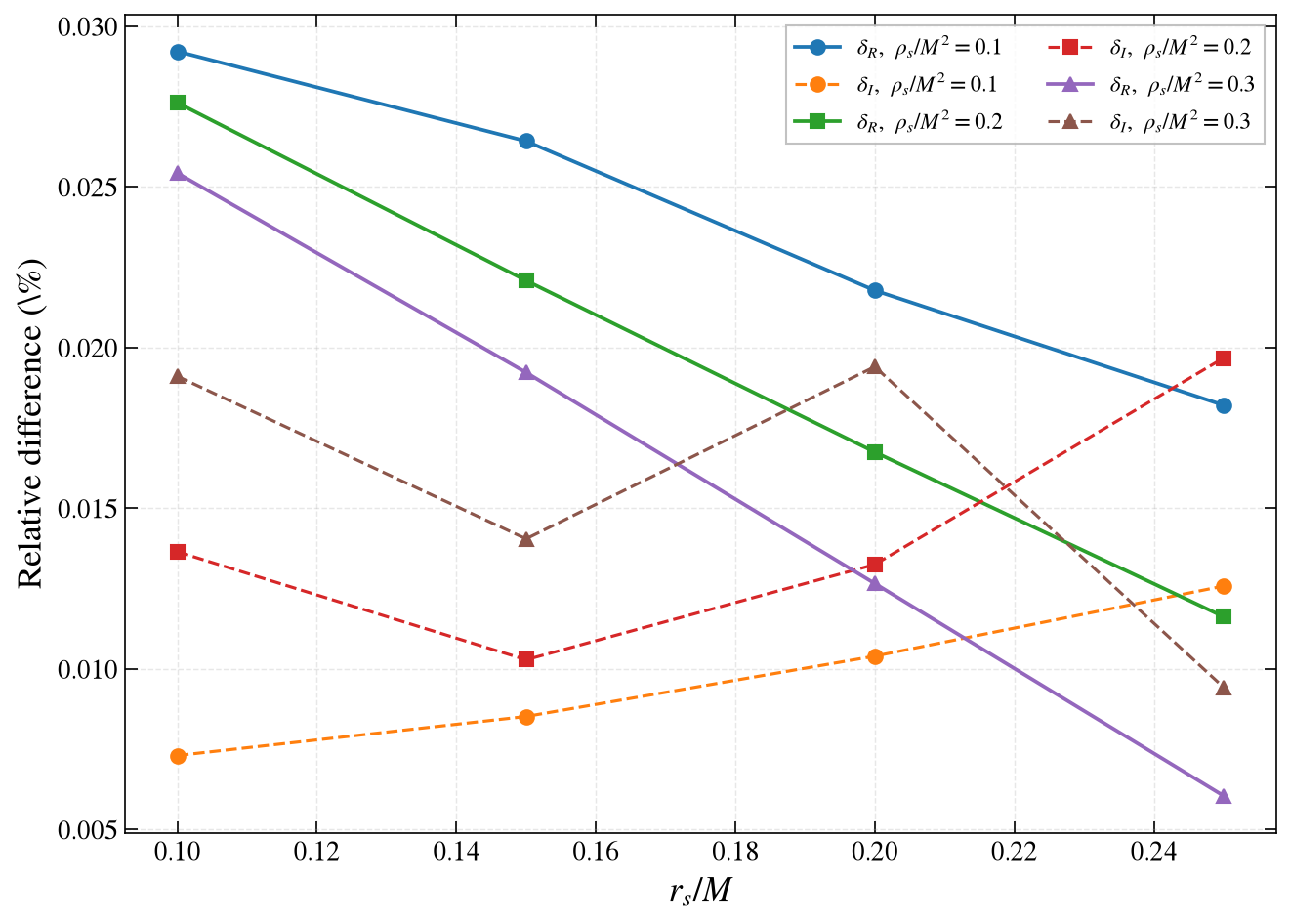}
    \caption{Relative differences between the Prony-extracted and sixth-order WKB QNM frequencies for the DC14 halo at $X=-2.6$ for  $r_s/M=0.1, 0.15, 0.2, 0.25$ and $\rho_s/M^2=0.1$, $0.2$, and $0.3$. }
    \label{fig:dc14_error_xm2p6}
\end{figure}
\begin{figure}[H]
    \centering
    \includegraphics[width=\columnwidth]
    {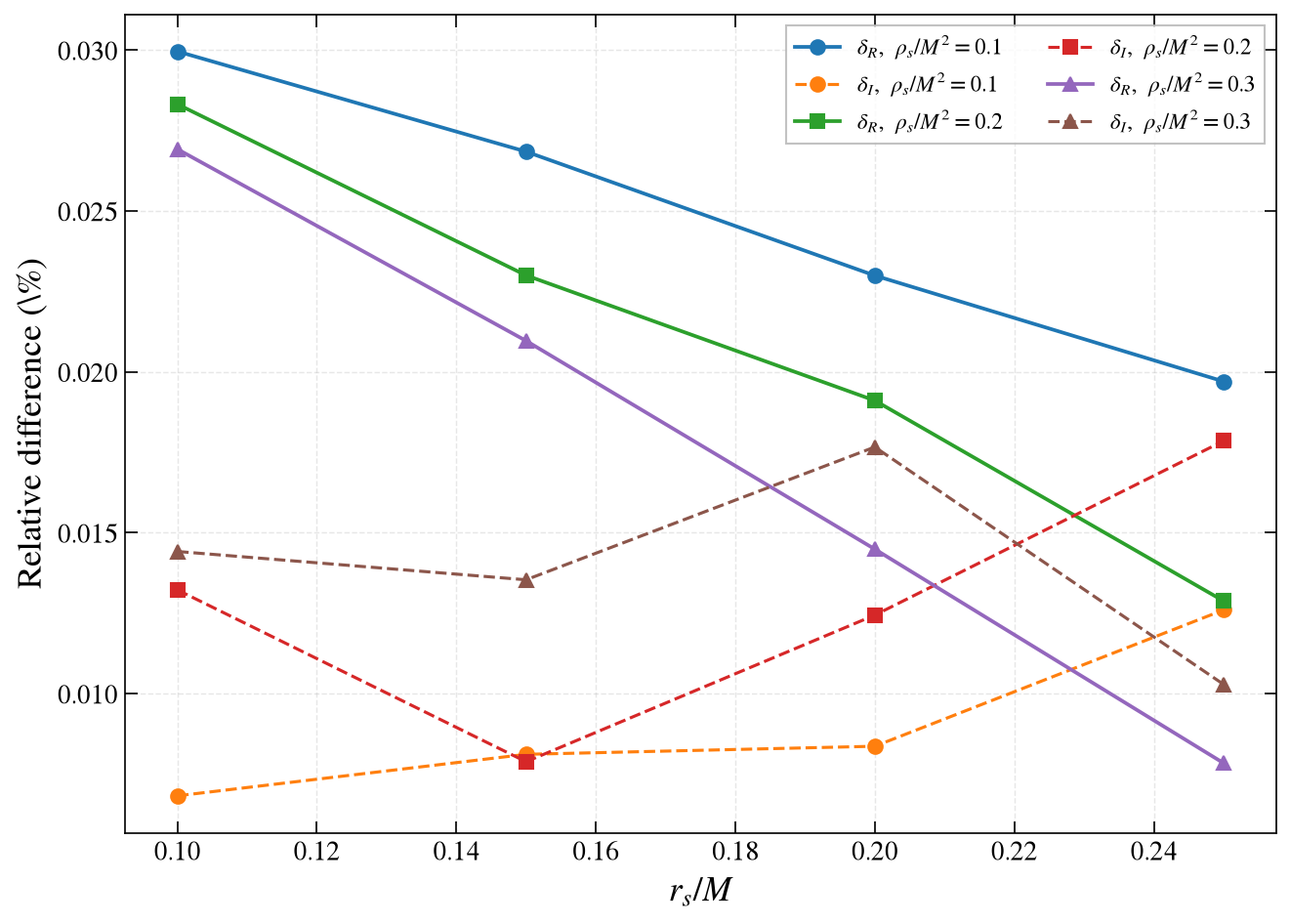}
    \caption{Relative differences between the Prony-extracted and sixth-order WKB QNM frequencies for the DC14 halo at $X=-2$ for  $r_s/M=0.1, 0.15, 0.2, 0.25$ and $\rho_s/M^2=0.1$, $0.2$, and $0.3$. }
    \label{fig:dc14_error_xm2p6}
\end{figure}

The QNM frequencies obtained from the Prony extraction show excellent agreement with the sixth-order WKB results across the full parameter range considered. The relative differences in both the real and imaginary parts remain very small, demonstrating the consistency of the two independent extraction methods. The corresponding
Prony-extracted and sixth-order WKB frequencies, together with their relative differences, are summarized in Tables~\ref{tab:prony_wkb_dc14_Xm4}--\ref{tab:prony_wkb_dc14_Xm2}

\section{Conclusion}
\label{sec:conclusion}

In this work, we developed and validated a numerical time-domain
pipeline for calculating the quasinormal modes of spherically
symmetric black holes in the presence of dark-matter environments.
The perturbation equation was evolved using a characteristic
finite-difference scheme in null coordinates, with the effective
potential evaluated as a function of the tortoise coordinate through
a numerical construction of the inverse mapping $r(r_*)$. The
resulting time-domain waveforms were analysed using Prony extraction
to obtain the complex QNM frequencies.

We applied the pipeline to black holes surrounded by two different
dark-matter halo profiles, namely the Dehnen and DC14 profiles, and
focused on the fundamental $\ell=2$, $s=2$ gravitational mode. By
varying the halo scale radius $r_s/M$, density normalization
$\rho_s/M^2$, and, for the DC14 profile, the parameter $X$, we find
systematic changes in both the oscillation frequency $\omega_R$ and
the damping rate $|\omega_I|$. The results demonstrate that the QNM
spectrum is sensitive to the structure and parameters of the
surrounding dark-matter distribution.

As an independent numerical cross-check, we compare the Prony-extracted frequencies with the sixth-order WKB approximation evaluated using the same effective potential. For the Dehnen profile, the relative differences remain
below $0.032\%$ in the real part and $0.023\%$ in the magnitude of
the imaginary part over the parameter range investigated. For the
DC14 profile, the corresponding relative differences remain below
$0.04\%$. The close agreement between the two independent approaches
provides a strong consistency check on both the numerical evolution
and the subsequent Prony extraction.

The results also show that the QNM shifts depend on the choice of
dark-matter profile, in addition to the halo parameters themselves. This highlights the importance of incorporating physically motivated environmental profiles when studying possible modifications of black hole ringdown signatures. The numerical pipeline developed here therefore provides a flexible framework for investigating QNMs in spherically symmetric black hole spacetimes surrounded by different matter distributions.

A natural extension of this work is to apply the same framework to additional dark-matter profiles and to investigate higher multipoles and overtones. In the present analysis, we have assumed $\delta T_{\mu\nu} = 0$, so that the dark-matter halo modifies the background geometry but does not contribute an explicit perturbation to the stress-energy tensor. Allowing for $\delta T_{\mu\nu} \neq 0$ would provide a further extension, enabling the study of coupled matter--gravitational perturbations and their impact on the QNM spectrum. Another important direction is to connect the
phenomenological halo parameters $r_s$ and $\rho_s$ considered here to physically motivated values inferred from galactic rotation-curve fits, as in Ref.~\cite{Zhang_2021}, in order to assess whether the resulting QNM shifts fall within the projected sensitivity of current and future gravitational-wave detectors such as LIGO, Virgo, and LISA. More broadly, the method developed here can be extended to study the resulting modifications to observable gravitational-wave ringdown signatures, providing a route towards assessing the prospects for constraining dark-matter environments with current and future gravitational-wave detectors.

\section*{Acknowledgements}
The author thanks Prof.~Anand Sen Gupta for his valuable comments and suggestions on this work, including ideas for extending the framework to observational gravitational-wave events.

The author acknowledges the use of artificial intelligence tools for
language editing, improving readability, informal discussion, and
literature-related checks during the preparation of this manuscript.
The author retains full responsibility for the scientific content,
analysis, results, and interpretation presented in this work.

\section*{Funding}
This research received no specific funding from any funding agency,
commercial, or not-for-profit organization.


\appendix
\section{Dehnen Halo Properties and Characteristic Orbital Radii}
\label{app:dehnen_properties}

\begin{figure}[htbp]
    \centering
    \includegraphics[width=1\linewidth]{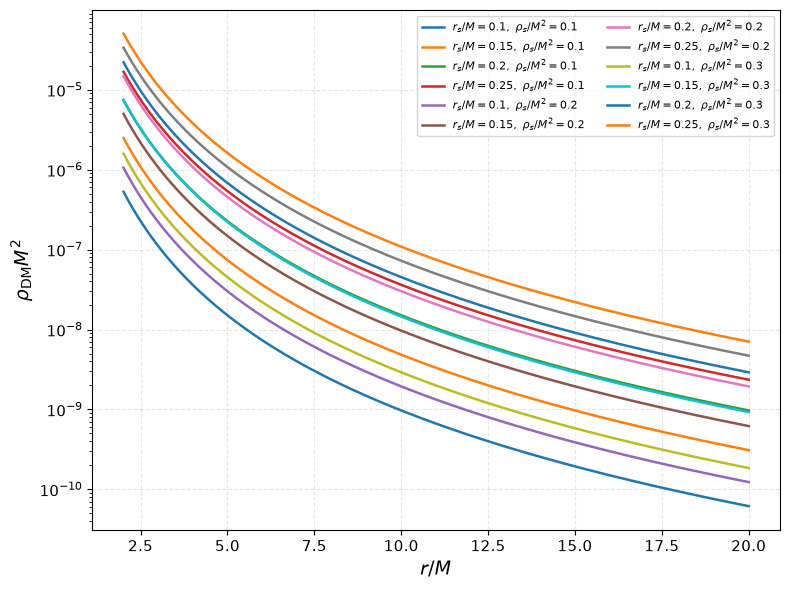}
    \caption{
    Dehnen dark-matter density profiles for representative values of
    the scale radius $r_s/M$ and density normalization $\rho_s/M^2$.
    }
    \label{fig:app_dehnen_density}
\end{figure}

\begin{figure}[htbp]
    \centering
    \includegraphics[width=1\linewidth]{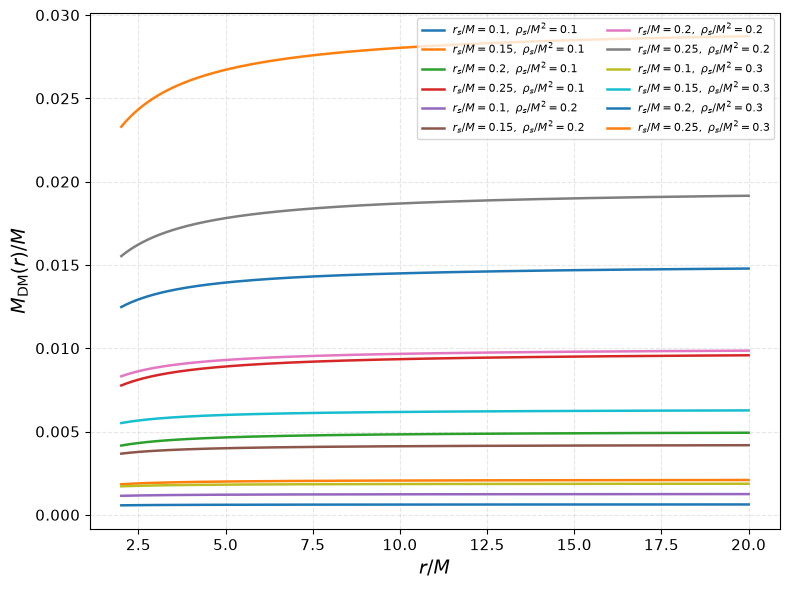}
    \caption{
    Enclosed dark-matter mass $M_{\rm DM}(r)$  for representative
    values of $r_s/M$ and $\rho_s/M^2$ considered in the analysis.
    }
    \label{fig:app_dehnen_mass}
\end{figure}

For completeness, we collect here the dark-matter density profile,
enclosed mass, and characteristic radii used in the numerical
calculation. The Dehnen density profile is written as
\begin{equation}
\rho_{\rm DM}(r)
=
\frac{\rho_s}
{\left(r/r_s\right)^\gamma
\left[1+\left(r/r_s\right)^\alpha\right]^{(\beta-\gamma)/\alpha}},
\label{eq:app_dehnen_density}
\end{equation}
where $r_s$ and $\rho_s$ denote the halo scale radius and density
normalization, respectively. The corresponding enclosed dark-matter
mass is
\begin{equation}
\begin{split}
M_{\rm DM}(r)
&=
\frac{4\pi\rho_s r_s^3}{3-\gamma}
\left(\frac{r}{r_s}\right)^{3-\gamma} \\
&\quad \times {}_2F_1\left[
\frac{3-\gamma}{\alpha},
\frac{\beta-\gamma}{\alpha};
1+\frac{3-\gamma}{\alpha};
-\left(\frac{r}{r_s}\right)^\alpha
\right].
\end{split}
\label{eq:app_dehnen_mass}
\end{equation}
The metric function used throughout the calculation is
\begin{equation}
f(r)=1-\frac{2[M+M_{\rm DM}(r)]}{r}.
\label{eq:app_metric}
\end{equation}
The corresponding density and enclosed-mass profiles for the
parameter range considered in this work are shown in
Figs.~\ref{fig:app_dehnen_density} and
\ref{fig:app_dehnen_mass}. These profiles provide the underlying
matter distribution entering the effective potential and therefore
the QNM calculation.

\vspace{10mm}

The characteristic radii associated with the modified geometry are
also evaluated. The event horizon is determined from $f(r_h)=0$,
while the photon-sphere radius satisfies
\begin{equation}
r_{\rm ph} f'(r_{\rm ph})-2f(r_{\rm ph})=0.
\label{eq:app_photon}
\end{equation}
The ISCO is obtained from
\begin{equation}
2[f'(r)]^2-f(r)f''(r)
-\frac{3f(r)f'(r)}{r}=0.
\label{eq:app_isco}
\end{equation}
For vanishing halo contribution these reduce to the familiar
Schwarzschild values $r_h/M=2$, $r_{\rm ph}/M=3$, and
$r_{\rm ISCO}/M=6$.

\begin{table}[htbp]
\centering
\caption{
Characteristic radii for the Dehnen dark-matter halo. The quantities
$r_h$, $r_{\rm ph}$, and $r_{\rm ISCO}$ denote the event-horizon,
photon-sphere, and ISCO radii, respectively.
}
\label{tab:app_dehnen_orbits}
\renewcommand{\arraystretch}{1.15}

\begin{tabular}{ccccc}
\toprule
$r_s/M$ & $\rho_s/M^2$
& $r_h/M$
& $r_{\rm ph}/M$
& $r_{\rm ISCO}/M$ \\
\midrule

0.10 & 0.10 & 2.00114 & 3.00172 & 6.00341 \\
0.15 & 0.10 & 2.00367 & 3.00558 & 6.01096 \\
0.20 & 0.10 & 2.00831 & 3.01271 & 6.02473 \\
0.25 & 0.10 & 2.01554 & 3.02384 & 6.04604 \\

\midrule

0.10 & 0.20 & 2.00228 & 3.00345 & 6.00682 \\
0.15 & 0.20 & 2.00734 & 3.01117 & 6.02192 \\
0.20 & 0.20 & 2.01664 & 3.02543 & 6.04952 \\
0.25 & 0.20 & 2.03113 & 3.04777 & 6.09226 \\

\midrule

0.10 & 0.30 & 2.00342 & 3.00518 & 6.01023 \\
0.15 & 0.30 & 2.01101 & 3.01677 & 6.03290 \\
0.20 & 0.30 & 2.02498 & 3.03818 & 6.07434 \\
0.25 & 0.30 & 2.04677 & 3.07176 & 6.13867 \\

\bottomrule
\end{tabular}
\end{table}


\section{DC14 Halo Properties and Characteristic Radii}
\label{app:dc14_properties}

The DC14 dark-matter halo profile used in this work is characterized
by three shape parameters, $\alpha$, $\beta$, and $\gamma$, which
depend on the stellar-to-halo mass ratio
\begin{equation}
X=\log_{10}\left(\frac{M_*}{M_{\rm halo}}\right).
\end{equation}
Following the DC14 parametrization, these quantities are given by
\begin{align}
\alpha &=
2.94-
\log_{10}\left[
10^{-1.08(X+2.33)}
+
10^{2.29(X+2.33)}
\right],
\label{eq:app_dc14_alpha}
\\
\beta &=
4.23+1.34X+0.26X^2,
\label{eq:app_dc14_beta}
\\
\gamma &=
-0.06+
\log_{10}\left[
10^{-0.68(X+2.56)}
+
10^{X+2.56}
\right].
\label{eq:app_dc14_gamma}
\end{align}

For the values of $X$ considered in this work, the resulting shape
parameters are
\begin{table}[htbp]
\centering
\caption{
DC14 shape parameters corresponding to the values of $X$ considered
in this work.
}
\label{tab:app_dc14_shape}
\renewcommand{\arraystretch}{1.15}
\begin{tabular}{cccc}
\toprule
$X$ & $\alpha$ & $\beta$ & $\gamma$ \\
\midrule
$-4.0$  & 1.13639 & 3.03000 & 0.92085 \\
$-3.0$  & 2.21400 & 2.55000 & 0.31193 \\
$-2.6$  & 2.59799 & 2.50360 & 0.23592 \\
$-2.0$  & 2.15198 & 2.59000 & 0.54712 \\
\bottomrule
\end{tabular}
\end{table}

The effective mass entering the metric is therefore
\begin{equation}
M_{\rm eff}(r)
=
M+M_{\rm DC14}(r),
\end{equation}
and the metric function becomes
\begin{equation}
f(r)
=
1-\frac{2M_{\rm eff}(r)}{r}.
\label{eq:app_dc14_metric}
\end{equation}

Since $M_{\rm DC14}(r)$ represents the enclosed halo mass, the effective
mass is radius dependent and increases as progressively more of the
halo is enclosed. This does not represent time-dependent accretion of
the black hole. For the halo profiles considered here,
$M_{\rm eff}(r)/r\rightarrow0$ at large radius, and hence
\begin{equation}
f(r)\rightarrow1,
\qquad r\rightarrow\infty,
\end{equation}
demonstrating the asymptotically flat behavior of the metric. At the
event horizon, $f(r_h)=0$.

\begin{figure}[htbp]
    \centering
    \includegraphics[width=1\linewidth]{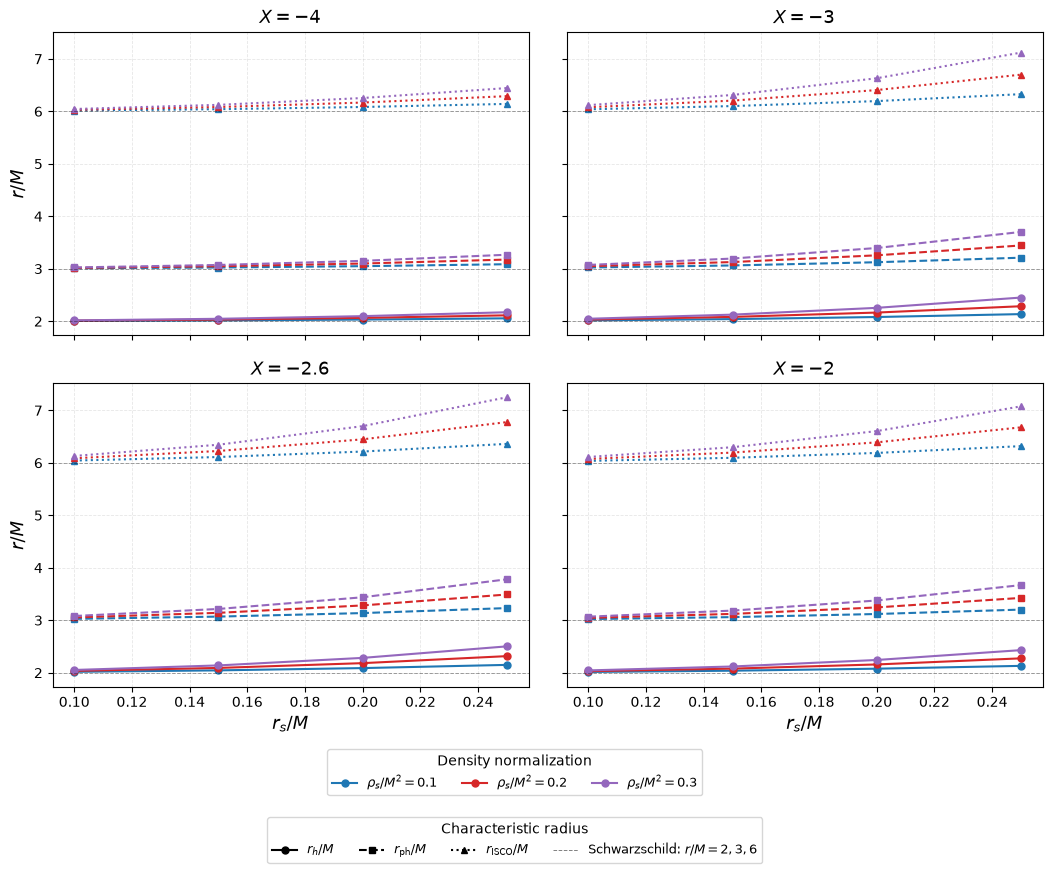}
    \caption{
    DC14 dark-matter density profiles for representative values of
    $X$, $r_s/M$, and $\rho_s/M^2$ considered in this work.
    }
    \label{fig:app_dc14_density}
\end{figure}

\begin{figure}[htbp]
    \centering
    \includegraphics[width=1\linewidth]{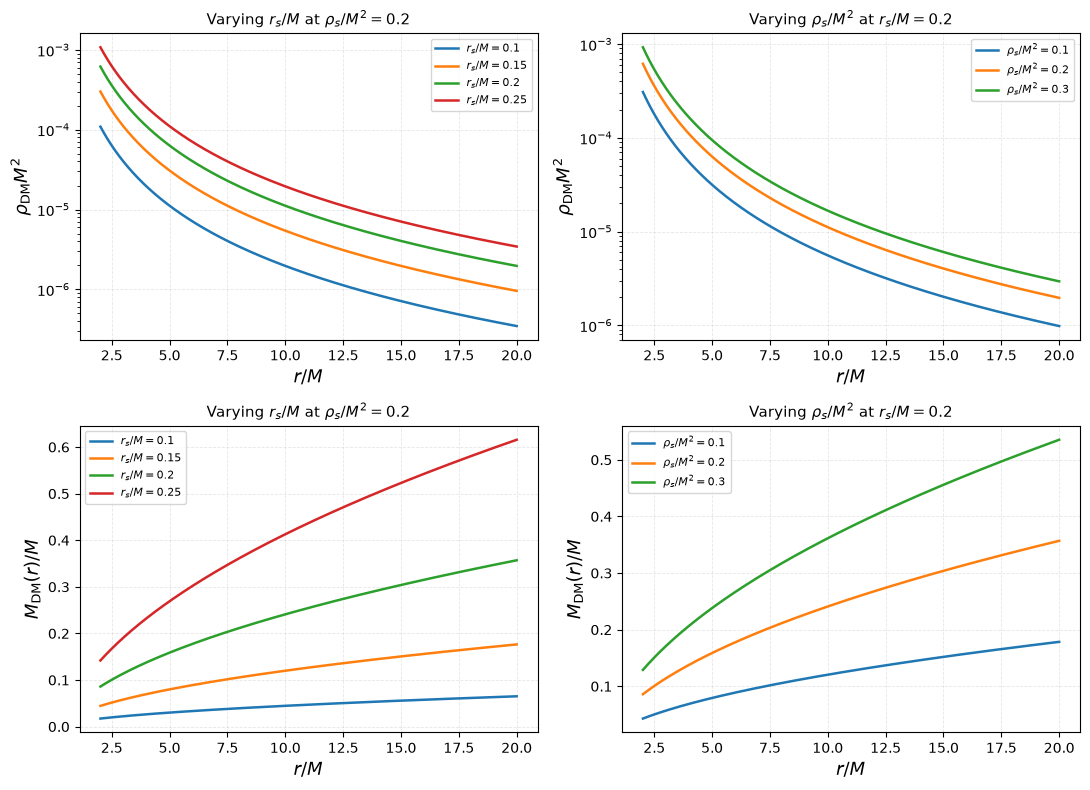}
    \caption{
    Enclosed dark-matter mass $M_{\rm DM}(r)/M$ ($M_{\rm DM}(r)$ $\equiv$ $M_{\rm DC14}(r)$) for $X=-2.6$. The increase with radius reflects the
    cumulative halo mass enclosed within radius $r$.
    }
    \label{fig:app_dc14_mass}
\end{figure}

The density and enclosed-mass profiles used in the numerical
calculation are shown in
Figs.~\ref{fig:app_dc14_density} and
\ref{fig:app_dc14_mass}. These plots illustrate the dependence of the
halo distribution on $r_s$, $\rho_s$, and the DC14 shape parameter
$X$.


Although the enclosed dark-matter mass $M_{\rm DC14}(r)$ increases
continuously with radius in Fig.~\ref{fig:app_dc14_mass}, the metric remains asymptotically flat.
For the DC14 profiles considered here, the mass contribution grows
more slowly than the radial coordinate at large $r$, such that
\begin{equation}
\frac{M_{\rm DC14}(r)}{r}\rightarrow 0,
\qquad r\rightarrow\infty.
\end{equation}
Consequently, the metric (lapse) function
\begin{equation}
f(r)
=
1-\frac{2\left[M+M_{\rm DC14}(r)\right]}{r}
\end{equation}
approaches
\begin{equation}
f(r)\rightarrow1,
\qquad r\rightarrow\infty.
\end{equation}
Thus, despite the continuous increase of the enclosed halo mass,
the spacetime approaches the asymptotically flat limit at large
radius.


The calculated characteristic radii for the four values of $X$ are
listed below.


\begin{table}[htbp]
\centering
\caption{Characteristic radii for the DC14 halo at $X=-4$.
}
\label{tab:app_dc14_radii_xm4}
\renewcommand{\arraystretch}{1.1}
\begin{tabular}{ccccc}
\toprule
$r_s/M$ & $\rho_s/M^2$ &
$r_h/M$ & $r_{\rm ph}/M$ & $r_{\rm ISCO}/M$ \\
\midrule
0.10 & 0.10 & 2.00542 & 3.00836 & 6.01467 \\
0.15 & 0.10 & 2.01539 & 3.02384 & 6.04079 \\
0.20 & 0.10 & 2.03182 & 3.04943 & 6.08264 \\
0.25 & 0.10 & 2.05541 & 3.08631 & 6.14136 \\
\cmidrule{1-5}
0.10 & 0.20 & 2.01085 & 3.01674 & 6.02939 \\
0.15 & 0.20 & 2.03090 & 3.04785 & 6.08197 \\
0.20 & 0.20 & 2.06416 & 3.09968 & 6.16720 \\
0.25 & 0.20 & 2.11255 & 3.17535 & 6.28917 \\
\cmidrule{1-5}
0.10 & 0.30 & 2.01629 & 3.02513 & 6.04414 \\
0.15 & 0.30 & 2.04651 & 3.07204 & 6.12355 \\
0.20 & 0.30 & 2.09702 & 3.15075 & 6.25369 \\
0.25 & 0.30 & 2.17143 & 3.26714 & 6.44337 \\
\bottomrule
\end{tabular}
\end{table}


\begin{table}[htbp]
\centering
\caption{Characteristic radii for the DC14 halo at $X=-3$.
}
\label{tab:app_dc14_radii_xm3}
\renewcommand{\arraystretch}{1.1}
\begin{tabular}{ccccc}
\toprule
$r_s/M$ & $\rho_s/M^2$ &
$r_h/M$ & $r_{\rm ph}/M$ & $r_{\rm ISCO}/M$ \\
\midrule
0.10 & 0.10 & 2.01558 & 3.02403 & 6.03883 \\
0.15 & 0.10 & 2.04081 & 3.06307 & 6.10024 \\
0.20 & 0.10 & 2.08034 & 3.12444 & 6.19507 \\
0.25 & 0.10 & 2.13585 & 3.21088 & 6.32706 \\
\cmidrule{1-5}
0.10 & 0.20 & 2.03132 & 3.04829 & 6.07814 \\
0.15 & 0.20 & 2.08274 & 3.12790 & 6.20406 \\
0.20 & 0.20 & 2.16532 & 3.25621 & 6.40503 \\
0.25 & 0.20 & 2.28580 & 3.44412 & 6.69923 \\
\cmidrule{1-5}
0.10 & 0.30 & 2.04720 & 3.07280 & 6.11794 \\
0.15 & 0.30 & 2.12581 & 3.19454 & 6.31152 \\
0.20 & 0.30 & 2.25518 & 3.39570 & 6.63036 \\
0.25 & 0.30 & 2.45117 & 3.70187 & 7.11938 \\
\bottomrule
\end{tabular}
\end{table}


\begin{table}[htbp]
\centering
\caption{Characteristic radii for the DC14 halo at  $X=-2.6$.
}
\label{tab:app_dc14_radii_xm2p6}
\renewcommand{\arraystretch}{1.1}
\begin{tabular}{ccccc}
\toprule
$r_s/M$ & $\rho_s/M^2$ &
$r_h/M$ & $r_{\rm ph}/M$ & $r_{\rm ISCO}/M$ \\
\midrule
0.10 & 0.10 & 2.01732 & 3.02668 & 6.04302 \\
0.15 & 0.10 & 2.04504 & 3.06952 & 6.11051 \\
0.20 & 0.10 & 2.08843 & 3.13676 & 6.21492 \\
0.25 & 0.10 & 2.14946 & 3.23163 & 6.36092 \\
\cmidrule{1-5}
0.10 & 0.20 & 2.03483 & 3.05366 & 6.08664 \\
0.15 & 0.20 & 2.09149 & 3.14125 & 6.22541 \\
0.20 & 0.20 & 2.18259 & 3.28257 & 6.44777 \\
0.25 & 0.20 & 2.31627 & 3.49074 & 6.77583 \\
\cmidrule{1-5}
0.10 & 0.30 & 2.05254 & 3.08095 & 6.13086 \\
0.15 & 0.30 & 2.13938 & 3.21525 & 6.34476 \\
0.20 & 0.30 & 2.28284 & 3.43801 & 6.69932 \\
0.25 & 0.30 & 2.50235 & 3.78042 & 7.24923 \\
\bottomrule
\end{tabular}
\end{table}


\begin{table}[htbp]
\centering
\caption{Characteristic radii for the DC14 halo at $X=-2$.
}
\label{tab:app_dc14_radii_xm2}
\renewcommand{\arraystretch}{1.1}
\begin{tabular}{ccccc}
\toprule
$r_s/M$ & $\rho_s/M^2$ &
$r_h/M$ & $r_{\rm ph}/M$ & $r_{\rm ISCO}/M$ \\
\midrule
0.10 & 0.10 & 2.01459 & 3.02249 & 6.03668 \\
0.15 & 0.10 & 2.03869 & 3.05981 & 6.09592 \\
0.20 & 0.10 & 2.07685 & 3.11906 & 6.18836 \\
0.25 & 0.10 & 2.13081 & 3.20309 & 6.31793 \\
\cmidrule{1-5}
0.10 & 0.20 & 2.02930 & 3.04518 & 6.07377 \\
0.15 & 0.20 & 2.07834 & 3.12112 & 6.19494 \\
0.20 & 0.20 & 2.15773 & 3.24447 & 6.38981 \\
0.25 & 0.20 & 2.27396 & 3.42576 & 6.67600 \\
\cmidrule{1-5}
0.10 & 0.30 & 2.04414 & 3.06807 & 6.11126 \\
0.15 & 0.30 & 2.11896 & 3.18396 & 6.29709 \\
0.20 & 0.30 & 2.24279 & 3.37651 & 6.60468 \\
0.25 & 0.30 & 2.43045 & 3.66958 & 7.07613 \\
\bottomrule
\end{tabular}
\end{table}


The characteristic radii show a systematic outward displacement
relative to the Schwarzschild values as the halo parameters are
increased. In particular, increasing either $r_s/M$ or
$\rho_s/M^2$ increases $r_h$, $r_{\rm ph}$, and $r_{\rm ISCO}$.
The magnitude of this displacement depends on the DC14 shape
parameter $X$, with the largest deviations in the parameter range
studied occurring for $X=-2.6$. These geometric changes are
consistent with the shifts observed in the QNM spectrum discussed
in the main text.

\section{Numerical Implementation and Validation}
\label{app:numerical}

This appendix summarizes the numerical implementation used for the
time-domain evolution and QNM extraction. The details are included
for reproducibility, while the main text focuses on the physical
results.


\subsection{Characteristic Evolution}
\label{app:characteristic}

The corresponding time-domain equation is
\begin{equation}
\frac{\partial^2 Z^{(-)}}{\partial t^2}
-
\frac{\partial^2 Z^{(-)}}{\partial r_*^2}
+
V^{(-)}(r)Z^{(-)}=0,
\label{eq:time_domain_wave}
\end{equation}
which we evolve in double-null coordinates,
\begin{equation}
u=t-r_*,
\qquad
v=t+r_*,
\end{equation}
for which
\begin{equation}
\frac{\partial^2 Z^{(-)}}{\partial u\,\partial v}
=
-\frac{1}{4}V^{(-)}(r)Z^{(-)}.
\label{eq:app_double_null}
\end{equation}

On the characteristic grid,
\begin{equation}
r_*=\frac{v-u}{2},
\end{equation}
and the effective potential is evaluated through the numerical
inversion $r=r(r_*)$. Thus,
\begin{equation}
V(r)\rightarrow
V\!\left[r\!\left(\frac{v-u}{2}\right)\right].
\end{equation}

\begin{figure}[htbp]
    \centering
    \includegraphics[width=0.82\linewidth]
    {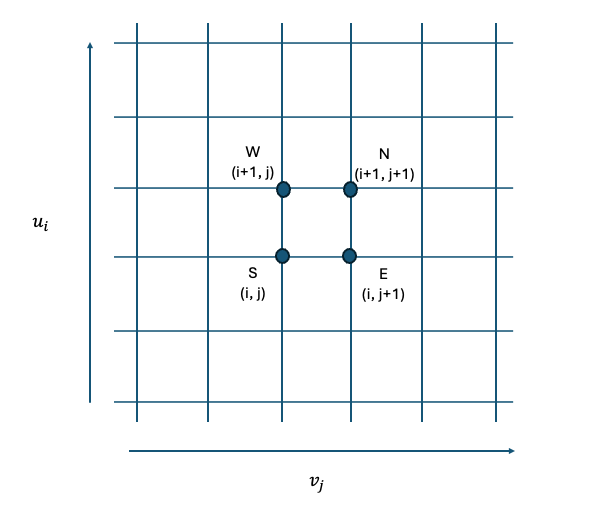}
    \caption{
    Characteristic integration stencil used in the numerical
    evolution. The future point $N$ is determined from the points
    $S$, $W$, and $E$.
    }
    \label{fig:app_grid_stencil}
\end{figure}

The field is discretized on a uniform grid with spacing $h$. For the
integration diamond shown in Fig.~\ref{fig:app_grid_stencil}, the
future point is obtained from the three previously known points as
\begin{equation}
\Psi_N
=
\Psi_W+\Psi_E-\Psi_S
-\frac{h^2}{8}
\left(V_W\Psi_W+V_E\Psi_E\right).
\label{eq:app_update}
\end{equation}
The discretization scheme to integrate Eq.~\ref{eq:app_update} was proposed by Gundlach, Price and Pullin~\cite{PhysRevD.49.883}.


\subsection{Initial Data and Waveform Extraction}
\label{app:initial_data}

A Gaussian perturbation is prescribed on the initial outgoing null
surface,
\begin{equation}
\Psi(u_0,v)
=
\exp\left[
-\frac{(v-v_c)^2}{2\sigma^2}
\right],
\end{equation}
while a zero-gradient condition is imposed on the other initial
null boundary,
\begin{equation}
\left.\partial_v\Psi\right|_{v=v_0}=0.
\end{equation}

The evolved solution is transformed to $(t,r_*)$ using
\begin{equation}
t=\frac{v+u}{2},
\qquad
r_*=\frac{v-u}{2}.
\end{equation}
A waveform at a fixed observation radius is therefore obtained by
sampling the characteristic grid along
\begin{equation}
v=u+2r_*^{\rm obs}.
\end{equation}

\begin{figure}[htbp]
    \centering
    \includegraphics[width=0.90\linewidth]
    {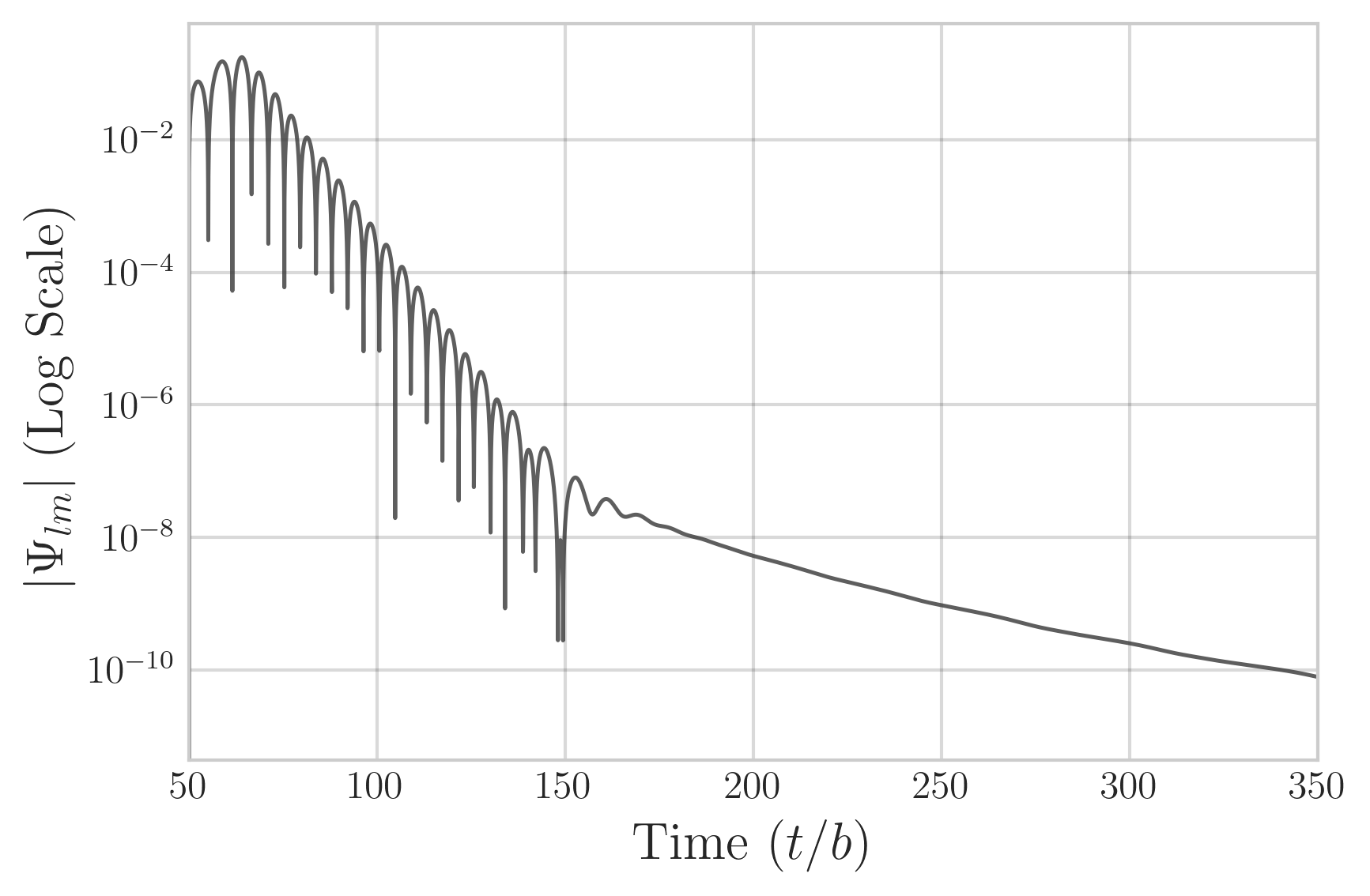}
    \caption{
    Representative time-domain waveform for $\ell=2$, showing the
    initial response, quasinormal ringing, and late-time tail.
    }
    \label{fig:app_tail}
\end{figure}

The initial prompt response is excluded from the QNM fit, and the
subsequent ringdown signal is used for frequency extraction. A
representative waveform, showing the transition from the initial
perturbation to the QNM ringing and late-time tail, is shown in
Fig.~\ref{fig:app_tail}.

\subsection{Prony Extraction}
\label{app:prony}

The ringdown signal is represented as a sum of damped complex exponentials~\cite{Chowdhury_2020, 1162433},
\begin{equation}
x[n]
\simeq
\sum_{j=1}^{p}C_j z_j^n,
\qquad
z_j=e^{-i\omega_j h}.
\end{equation}

The corresponding linear prediction relation is
\begin{equation}
\sum_{m=0}^{p}\alpha_m x[n-m]=0,
\qquad
\alpha_0=1.
\end{equation}

The prediction coefficients are obtained from an overdetermined
linear system using singular-value decomposition. The roots of
\begin{equation}
\mathcal{P}(z)
=
1+\sum_{m=1}^{p}\alpha_m z^{-m}
\end{equation}
then give the complex QNM frequencies through
\begin{equation}
\omega_j=\frac{i}{h}\ln z_j.
\end{equation}

The mode amplitudes are subsequently determined by linear
least-squares fitting, and the stable dominant mode is identified as
the fundamental QNM.


\subsection{Numerical Convergence and Stability}
\label{app:convergence}

The numerical accuracy of the evolution is tested using three grid
resolutions, $h$, $h/2$, and $h/4$. The convergence factor is defined
as
\begin{equation}
Q(r_*)
=
\frac{
|\Psi_h(r_*)-\Psi_{h/2}(r_*)|
}{
|\Psi_{h/2}(r_*)-\Psi_{h/4}(r_*)|
}.
\end{equation}

\begin{figure}[htbp]
    \centering
    \includegraphics[width=0.92\linewidth]{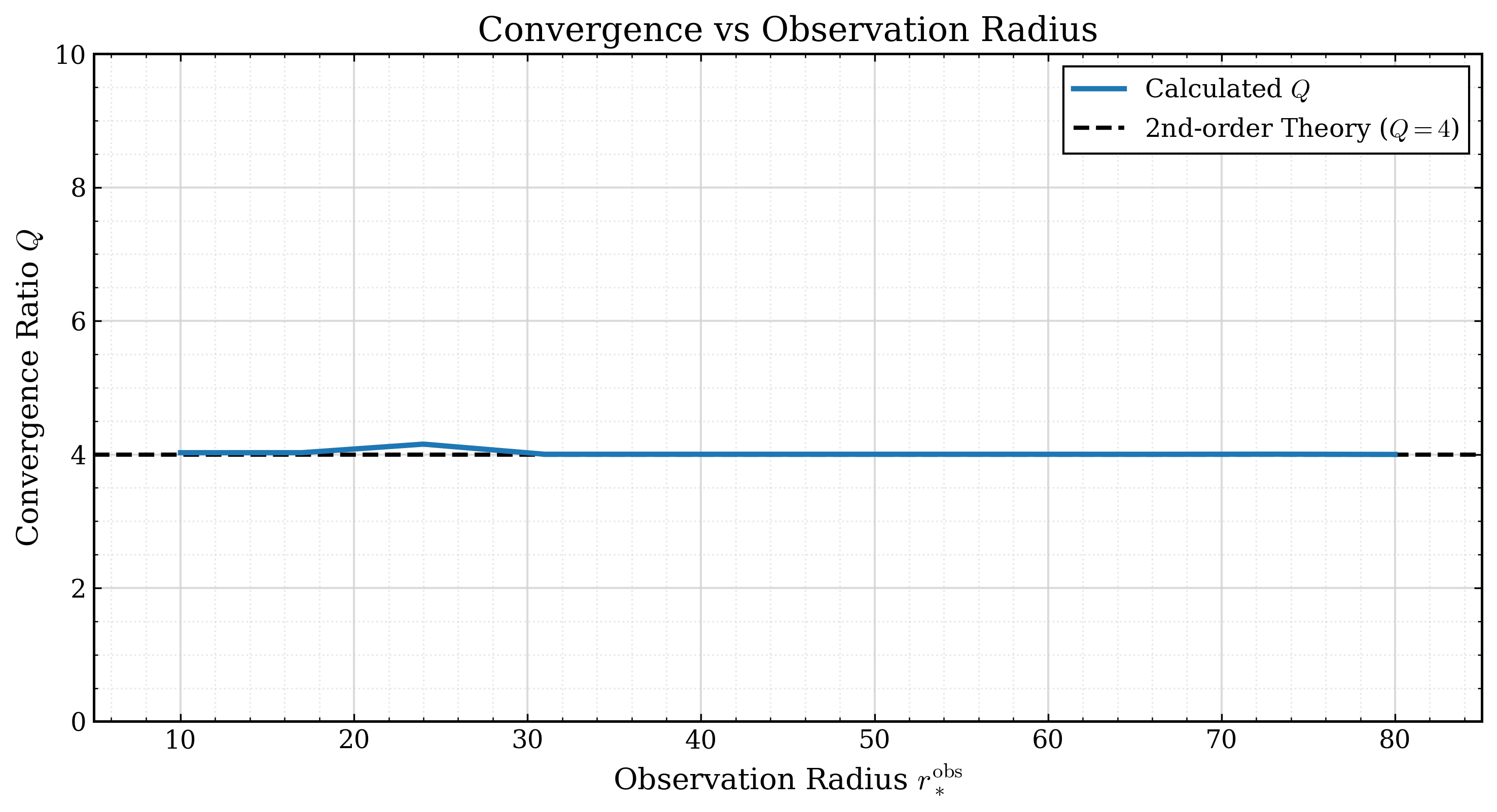}
    \caption{
    Convergence factor $Q$ obtained from simulations with resolutions
    $h$, $h/2$, and $h/4$. The result approaches the expected value
    $Q=4$ for a second-order convergent scheme.
    }
    \label{fig:app_convergence}
\end{figure}

For the second-order scheme used here, the expected value is
$Q\rightarrow4$. The numerical result, shown in
Fig.~\ref{fig:app_convergence}, approaches this value across the
computational domain.

\clearpage
\bibliography{ref}

\end{document}